\documentclass[pre,aps,showpacs,twocolumn]{revtex4}
\usepackage{amssymb,bm}
\usepackage[dvips]{graphicx}
\usepackage[dvips]{color}

\begin{document}

\title{Nearly spherical vesicles in an external flow}

\author{V. V. Lebedev, K. S. Turitsyn, and S. S. Vergeles}
\affiliation{Landau Institute for Theoretical Physics,
 Moscow, Kosygina 2, 119334, Russia}
\date{\today}

 \begin{abstract}

Tank-treading, tumbling and trembling are different types of the vesicle behavior in an
external flow. We derive a dynamical equation for nearly spherical vesicles enabling to
establish a phase diagram of the system predicting the regimes. The diagram is drawn in
terms of two dimensionless parameters depending on the vesicle excess area, fluid
viscosities, membrane viscosity and bending modulus, strength of the flow, and ratio of
the elongational and rotational components of the flow. The tank-treading to tumbling
transition occurs via a saddle-node bifurcation whereas the tank-treading to trembling
transition occurs via a Hopf bifurcation. We establish a critical slowing near the
merging point of the transition lines.

 \end{abstract}


\maketitle

\section{Introduction}
\label{sec:intro}

\begin{figure}[t]
 \includegraphics[width=3.5in,angle=0]{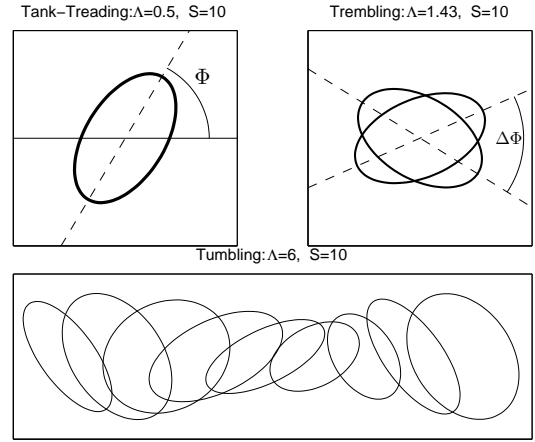}
 \caption{Vesicle projections to the shear plane in
 the tank-treading, trembling and tumbling regimes.}
 \label{fig:sequence}
 \end{figure}

Vesicles, which are closed membranes separating two regions filled up by generally
different fluids, have very much in common with biological cells and have a number of
applications in pharmaceutics. The truly non-equilibrium problem of describing the
vesicle dynamics in external flows attracted lots of attention both from experimentalists
\cite{HBVD97,SBAM98,ALV02,AV05,KS05,MVAV06,KS06} and theoreticians
\cite{82KS,KWSL96,S99,O00,SS01,NT02,BM02,BM03,RBM04,NG04,BRSBM04,BKM05,
NG05_1,NG05_2,NG05_3,M06,VG07,NG07}. In experiments different regimes of vesicle
dynamics were observed, usually in share flows. In the tank-treading regime a vesicle
shape is stationary. In the tumbling regime a vesicle experiences periodic flipping in
the shear plane. Trembling, experimentally discovered in the work \cite{KS06}, is an
intermediate regime between tank-treading and tumbling, in the regime a vesicle trembles
around the flow direction. Theoretical studies \cite{M06} and \cite{NG07} have
predicted similar vesicle behavior and called it vacillating-breathing and swinging
respectively. The different types of the motion are illustrated in Fig.
\ref{fig:sequence}.

In this paper we focus on constructing the phase diagram, which would predict what kind
of regime corresponds to a given set of external parameters, such as vesicle
non-sphericity, viscosity contrast, strength of external flow. Although this problem was
addressed previously by different authors, it is still far from being closed. As long as
no analytic solution of this problem exists, theoretical studies were based either on
numerical simulations or on some approximations allowing analytical treatment. Numerical
investigations of this problem involved several different computational schemes,
including boundary element method \cite{KWSL96,SS01}, mesoscopic particle-based
approximation \cite{NT02,NG04,NG05_1,NG05_2,NG05_3}, and an advected field approach
\cite{BM02,BM03,BRSBM04,BKM05}. These approaches have shown qualitative agreement with
experiment. However, they did not solve the problem of constructing the vesicle phase
diagram completely. Analytical studies of the problem can be divided in two major
classes. In the first one, see Refs. \cite{RBM04,NG04,NG07}, phenomenological models of
vesicle dynamics based on the classical work of Keller and Skallak \cite{82KS} were
proposed and proved themselves to be rather efficient in explaining particular
experiments. However, they cannot cover wide range of parameters and cannot,
consequently, be used for constructing a phase diagram for the vesicle dynamics. In the
second series of works, Refs. \cite{S99,O00,M06,VG07}, the studies are focused on
nearly spherical (quasi-spherical) vesicles whose shape is close to spherical one and can
be parameterized by an expansion over spherical harmonics.

We propose a natural extension of the theory developed for quasi-spherical vesicles by
accounting higher-order expansion terms. A perturbation scheme around the Lamb solution
for spherical body in an external flow \cite{32Lamb} allows one to derive the dynamic
equations for the shape and the orientation of a vesicle and to investigate them
analytically. We show that the high-order in the vesicle distortion terms produce a
qualitative change in the phase diagram and make it significantly more complicated. The
resulting diagram contains all three types of vesicle behavior which were observed
experimentally. In our work, we analyze how the vesicle dynamics depends on different
control parameters, such as viscosity contrast, vesicle excess area, internal membrane
viscosity, strength of the flow, and ratio of the elongational and rotational components
of the flow. We analyze also the vesicle orientation in the tank-treading regime. Some of
the results, derived in this paper were already reported in \cite{LTV07}. Here we present
significantly more detailed derivations and analyze several aspects of the problem which
were not discussed in \cite{LTV07}.

Speaking about membranes we have in mind lipid bilayers, the simplest type of the
biological membranes. Physical properties of such objects have been extensively studied,
both experimentally and theoretically (see e.g. the books \cite{ML87,SC87,NP89} and the
reviews \cite{BP84,PA91,PO92}). There are several features of the membranes which are
important for our analysis. First, we assume that the membrane is in a fluid (isotropic)
state (is a $2d$ liquid), which is typical of lipid bilayers under normal conditions.
Second, we assume that the vesicle has an excess area, that enables one to treat the
membrane as incompressible. Third, we assume that the membrane is impermeable to the
surrounding fluids, the condition is usually well satisfied in experiment. Finally, we
take into account the membrane internal viscosity, which can play an essential role, say,
in the vicinity of the lipid-bilayer melting point \cite{DPD00}.

The structure of our paper is as follows. In Section \ref{sec:basic} we expose basic
theoretical facts concerning physics of membranes. A special attention is paid to dynamic
properties of the membranes which can be treated as moving interfaces immersed in a $3d$
fluid. In Section \ref{sec:nearly} we formulate peculiarities of nearly spherical
vesicles, analyze their equilibrium properties, and derive phenomenological equation
which describe their dynamics in weak external flows. In Section \ref{sec:perturb} we
establish a dynamic equation for the vesicle evolution obtained in the framework of
expansion over deviations from the ideal spherical shape. We introduce dimensionless
parameters controlling the vesicle behavior. In Section \ref{sec:planar} we consider the
case of planar external velocity field. In this case the dynamic equations can be
essentially simplified enabling one to establish the phase diagram of the system
containing domains corresponding to tank-treading, trembling and tumbling. In Section
\ref{sec:limit} we analyze limit cases of weak and strong external flows where it is
possible to study the vesicle behavior in details. In Conclusion we discuss some outcomes
of our work and its possible extensions. Some technical details of the perturbation
expansion are presented in Appendix.

\section{Basic Relations}
\label{sec:basic}

We are interested in processes which take place at scales of the order of the vesicle
size which are assumed to be much larger than the membrane thickness. This assumption is
well justified for giant vesicles, usually examined in experiment. In the main
approximation, the membrane can be treated as infinitesimally thin, that is as a $2d$
object (film) immersed into a $3d$ fluid. In this case the vesicle is characterized by
its geometrical shape. In other words, in this limit the vesicles can be considered as
active interfaces separating different pieces of the fluid.

The two membrane properties, its incompressibility and impermeability, imply that both
the vesicle volume ${\cal V}$ and its surface area ${\cal A}$ are conserved, provided the
vesicle has an excess area. The latter can be characterized by a dimensionless factor
$\Delta$, which is traditionally introduced as
 \begin{equation}
 \label{Delta}
    {\cal A}=(4\pi+\Delta){r}_0^2, \qquad
    {\cal V} = 4\pi {r}_0^3/3 \,,
 \end{equation}
where $r_0$ is a vesicle ``radius'' determined by its volume. The excess area factor is
non-negative, $\Delta\geq0$, and the minimal value $\Delta =0$ corresponds to the ideal
spherical geometry. The nearly spherical (quasi-spherical) vesicles are characterized by
the condition $\Delta\ll1$.

The energy of the membrane is determined by its bending distortions and can be written as
the following surface integral \cite{70Can,73Hel,74Eva,75Hel}
 \begin{equation}
 {\cal F}^{({\mathrm b})} =\int dA\
 \left({\kappa}H^2/2
 +{\bar\kappa}K\right) \,,
 \label{Helfrich}
 \end{equation}
over the membrane position. Here $\kappa$ and $\bar\kappa$ are bending modules, $H$ and
$K$ are the mean and Gaussian curvatures, respectively. They are related to the local
curvature radii of the membrane $R_1$ and $R_2$ as
 \begin{equation}
 H=R_1^{-1}+R_2^{-1}\,, \quad
 K=R_1^{-1}R_2^{-1}\,.
 \label{curv}
 \end{equation}
In accordance with the Gauss-Bonnet theorem, the second term in the right-hand side of
Eq. (\ref{Helfrich}) (with the factor $\bar\kappa$) is invariant under smooth
deformations of the membrane shape. Therefore the term is irrelevant provided the vesicle
topology is fixed.

Besides the bending energy (\ref{Helfrich}) the membrane is characterized by its surface
tension $\sigma$. One should stress that for vesicles with excess area the surface
tension is an auxiliary variable adjusting to other quantities characterizing the
membrane and the surrounding flow to ensure the membrane incompressibility.

\subsection{Flow near the vesicle}

In this paper we consider the situation where both fluids, outside and inside the
vesicle, are Newtonian. Furthermore, we assume that the Reynolds number associated with
the vesicle dynamics is vanishingly small, which is the case in real experiments
\cite{HBVD97,SBAM98,ALV02,AV05,KS05,MVAV06,KS06}. Under these assumptions the fluids can
be described by the Stokes equation
 \begin{equation}
 \varrho\,\partial_t \bm{v}
 = \eta \nabla^2 \bm{v} -\nabla P \,,
 \label{Stokes}
 \end{equation}
where $P$ is pressure, $\bm v$ is the fluid velocity, $\varrho$ is its mass density, and
$\eta$ is its dynamic viscosity. The equation (\ref{Stokes}) has to be supplemented by
the incompressibility condition $\nabla\bm v=0$, which leads to the Laplace equation
$\nabla^2 P=0$ for the pressure.

We divide the flow near the vesicle into two parts: an external flow, which would be
observed in the fluid in the absence of the vesicle, and an induced flow, which is
excited as a result of the vesicle reaction to the external flow. The external flow is
assumed to be stationary, its velocity is designated as $\bm V$. One should remember that
the vesicle is advected by the flow, and therefore the above assumption should be valid
in the Lagrangian reference frame attached to the vesicle. Below, we neglect the term
with the time derivative in Eq. (\ref{Stokes}) since the characteristic time scale
associated with the vesicle dynamics is assumed to be large compared to the viscous
relaxation time $\varrho r_0^2/\eta$.

We assume that the characteristic spatial scale of the external flow is much larger than
the vesicle size. In this case the external velocity $\bm V$ near the vesicle can be
approximated by a linear profile, determined by a derivative matrix $\partial_k V_i$. The
incompressibility condition implies that the matrix $\partial_k V_i$ is traceless.
Generally, the external flow has two contributions, elongational and rotational:
 \begin{equation}
 \partial_k V_i= s_{ik} - \epsilon_{ikj} \omega_j \,,
 \label{graddec}
 \end{equation}
where $\hat s$ is the strain matrix (symmetric part of the matrix $\partial_k V_i$) and
$\bm\omega$ is the angular velocity vector. The strain can be characterized by its
strength $s$, defined as $s^2=(1/2) \mathrm{tr}\ \hat s^2$. Note that for a shear flow
$s=|\bm\omega|=\dot\gamma/2$, where $\dot\gamma$ is the shear rate.

The fluids inside and outside the vesicle are assumed to be different. We use the
designations $\eta$ for the external fluid viscosity, the viscosity of the internal fluid
is designated as $\tilde\eta$. An important parameter which controls the tank-treading to
tumbling transition is the viscosity contrast $\tilde\eta/\eta$. The limit where the
viscosity contrast tends to infinity corresponds to a solid body behavior of the vesicle,
that was concerned by Jeffery \cite{jeffery}.

The membrane moves together with the fluid that is the velocity field $\bm v$ is
continuous on the membrane and the field $\bm v$ determines the membrane velocity as well
as the fluid velocity. For relatively slow processes, we are investigating, the membrane
can be treated as locally incompressible, that leads to the condition
 \begin{equation}
 \partial^\perp_i v_i=0, \qquad \mathrm{where} \quad
 \partial^\perp_i=\delta^\perp_{ik}\partial_k \,,
 \label{sincomp}
 \end{equation}
to be satisfied on the membrane. Here $\delta^\perp_{ik}$ is the projector to the
membrane, it can be written as $\delta^\perp_{ik}=\delta_{ik}-l_i l_k$ where $\bm l$ is
the unit vector normal to the membrane. The $3d$ incompressibility condition $\nabla\bm
v=0$ together with Eq. (\ref{sincomp}) leads to the relation $l_il_k\partial_i v_k=0$, to
be satisfied at both sides of the membrane.

\subsection{Membrane stress}

The membrane reaction is characterized by its surface stress tensor $T^{(\mathrm
s)}_{ik}$. There are three contributions to the tensor related to the bending energy
(\ref{Helfrich}), to the surface tension of the membrane, and to the internal membrane
viscosity:
 \begin{equation}
 T^{(\mathrm s)}_{ik}=T^{(\kappa)}_{ ik}
 -\sigma\delta^{\perp}_{ik}
 -\zeta \delta^{\perp}_{ij}\delta^{\perp}_{kn}
 (\partial_j v_n +\partial_n v_j)  \,,
 \label{me4}
 \end{equation}
where $\sigma$ is the surface tension coefficient and $\zeta$ is the membrane ($2d$)
dynamic viscosity. Note that the surface tension $\sigma$ plays an auxiliary role being
adjusted to other stresses to ensure the local membrane incompressibility. An expression
for the bending contribution to the surface stress tensor was found in the work
\cite{89LM} (see also the book \cite{93KL}). It can be written as
 \begin{equation}
 T^{(\kappa)}_{ik}=
 \kappa\left(-\frac{1}{2}H^2\delta^{\perp}_{ik}
 + H \partial^\perp_i l_k
 -l_i \partial^\perp_k H \right),
 \label{kap}
 \end{equation}
where $H=\nabla \bm l$ is the membrane mean curvature and $\partial^\perp_i$ is defined
by Eq. (\ref{sincomp}).

The surface force $\bm f$ (force per unit area) associated with the membrane stress
tensor $T^{(\mathrm s)}_{ik}$ can be calculated as $f_i=-\partial_k^\perp T^{(\mathrm
s)}_{ik}$. Then one obtains from the expressions (\ref{me4},\ref{kap}) three
contributions to the surface force
 \begin{equation}
 \bm f=\bm f^{(\kappa)}+\bm f^{(\sigma)}
 + \bm f^{(\mathrm v)}\,,
 \label{surforce}
 \end{equation}
where
 \begin{eqnarray}
 f^{(\kappa)}_i =\kappa\left[H\left({H^2}/{2}-2K\right)
 +\Delta^\perp H\right] l_i \,,
 \label{Hforce} \\
 f^{(\sigma)}_i =-H\sigma l_i
 +\partial^\perp_i\sigma \,, \hspace{3cm}
 \label{tforce} \\
 f^{(\mathrm v)}_i=\zeta\left[\delta^\perp_{ij}
 \Delta^\perp v_j-H l_n \partial^\perp_i v_n
 -2l_i(\partial^\perp_n l_j)\partial^\perp_j v_n\right].
 \label{vforce} \end{eqnarray}
Here, again, $H$ and $K$ are the mean curvature and the Gaussian curvature of the
membrane, and $\Delta^\perp$ is the Laplace-Beltrami operator, $\Delta^\perp
=\partial^\perp_i\partial^\perp_i$, associated with the mebrane. Note that the expression
for the force (\ref{Hforce}) can also be derived by calculating the variation of the
bending energy (\ref{Helfrich}) due to infinitesimal membrane deformations \cite{89Hel}.

The surface force $\bm f$ is compensated by the momentum flux from the surrounding medium
to the membrane. This flux consists of two parts, related to the fluid pressure and the
fluid viscosity. As a result of the balance, we find the following relations
 \begin{eqnarray}
 -\kappa[H(H^2/2-2K) +\Delta^\perp H]+\sigma H
 \nonumber \\
 +2\zeta \partial^\perp_i l_n\partial^\perp_n v_i
 =P_{in}-P_{out},
 \label{normal} \\
 \partial^\perp_i \sigma
 +\zeta(\delta^\perp_{ij} \Delta^\perp v_j
 -Hl_n \partial^\perp _iv_n)
 \nonumber \\
 = l_k\,
 [\tilde\eta(\partial_i v_k+\partial_k v_i)_{in}
 -\eta(\partial_i v_k+\partial_k v_i)_{out}],
 \label{tangent}
 \end{eqnarray}
for the normal and tangential to the membrane components of the force. Here, we assumed
that the unit vector $\bm l$ is directed outwards the vesicle and the subscripts $in$ and
$out$ label regions inside and outside the vesicle, respectively. Thus, $P_{in}-P_{out}$
is the pressure difference between the inner and outer regions that is the pressure jump
on the membrane. Note that a fluid viscous contribution is absent in Eq. (\ref{normal})
due to the condition $l_il_j\partial_i v_j=0$, following from the membrane
incompressibility (see above).

To find the velocity field at a given membrane shape one should solve the stationary
Stokes equation $\eta\nabla^2 \bm v =\nabla P$ (inside and outside the vesicle) with the
boundary conditions (\ref{sincomp},\ref{normal},\ref{tangent}) on the membrane. An
additional boundary condition reads that $\bm v\to\bm V$ far away from the membrane. Note
that due to linearity of the equations and the boundary conditions for the velocity a
solution of the equations can be written as a sum
 \begin{equation}
 \bm v=\bm v^{(s)}+\bm v^{(\kappa)},
 \label{decompos}
 \end{equation}
where $\bm v^{(s)}$ is proportional to the gradient of the external flow (\ref{graddec}),
and $\bm v^{(\kappa)}$ is proportional to the bending modulus $\kappa$. Of course, $\bm
v^{(s)}$ and $\bm v^{(\kappa)}$ are complicated functions of the vesicle shape.

\subsection{Membrane shape parametrization}
\label{subsec:parametr}

Below, we use a particular parametrization of the vesicle shape which is
  \begin{equation}
  \label{function_u}
  r = r_0[1 + u(\theta,\varphi)] \,,
  \end{equation}
where $r_0$ is determined by the relation (\ref{Delta}). We use the spherical coordinate
system with variables $r,\theta,\phi$ and with the origin in the center of the vesicle.
The dimensionless radial displacement $u$ characterizes the deviations of the membrane
shape from the spherical one.

There are constraints imposed on the function $u(\theta,\varphi)$ due to the conditions
(\ref{Delta}). In terms of the displacement $u$ the conditions can be rewritten as
 \begin{eqnarray}
 \int {d}\varphi\, d\theta\, \sin\theta\,
 (u+u^2+u^3/3) =0 \,,
 \label{constraint} \\
 \Delta= \int {d}\varphi\, d\theta\, \sin\theta\,
 \Big\{ (1+u)
 \hspace{1cm}\label{constrainta}
 \\ \nonumber
 \left[
 (1+u)^2+{(\partial u/\partial\theta)^2
       +\sin^{-2}\theta (\partial u/\partial\varphi)^2}
 \right]^{1/2}-1\Big\} \,.
 \end{eqnarray}
The relations (\ref{constraint},\ref{constrainta}) are formally exact. However, they can
be directly used only if $u(\theta,\varphi)$ is a single-valued function.

Advection of the membrane by the surrounding fluid implies the following kinematic
relation
 \begin{equation}
 \partial_t u =\frac{1}{r_0}v_r-
 \frac{1}{r}\left(v_\theta \partial_\theta u
 +\frac{1}{\sin\theta}v_\varphi \partial_\varphi u\right).
 \label{kinema}
 \end{equation}
Here $v_r,v_\theta,v_\varphi$ are spherical components of the velocity $\bm v$ taken at
the membrane, that is at $r$ determined by Eq. (\ref{function_u}). Again, the relation
(\ref{kinema}) is formally exact, but can be directly used only if $u(\theta,\varphi)$ is
a single-valued function.

Closed equation which describes the dynamics of the membrane displacement $u$ can be
derived in two steps. First, one should find the fluid velocity profile for a given
displacement $u(\theta,\varphi)$. Second, one should use the kinematic relation
(\ref{kinema}). This procedure results in a closed non-linear equation for $u$. Note that
due to the property (\ref{decompos}) the expression is a sum of two terms, which are
proportional to the external flow gradient (\ref{graddec}) and to the bending modulus
$\kappa$. Of course, both terms are non-linear in $u$.

\section{Perturbation expansion: weak flows}
\label{sec:nearly}

Below, we consider nearly spherical (quasispherical) vesicles, that is the excess area
parameter $\Delta$, introduced by Eq. (\ref{Delta}), is considered to be small. In this
case the dimensionless displacement $u$ is small and it is possible to develop an
expansion over $u$. This perturbation series is a basis for subsequent consideration.

It is natural to represent the function $u(\theta,\varphi)$ as a sum over spherical
harmonics:
 \begin{equation}
 u = \sum_{l,m} u_{l,m} {\cal Y}_{l,m}(\theta,\phi).
 \label{harmonics}
 \end{equation}
The homogeneous contribution to $u$ (its zero angular harmonic $u_{0,0}$) can be
expressed via the inhomogeneous one (non-zero harmonics) from the relation
(\ref{constraint}) which reflects the volume conservation. Substituting the obtained
expression for the zero angular harmonic into Eq. (\ref{constrainta}) we obtain an
expression for $\Delta$ whose expansion over $u$ starts from the second order term.
Therefore the displacement $u$ can be estimated as $\sqrt\Delta$. That justifies the
expansion over $u$.

Different angular harmonics in $u$ play different roles. The zero harmonic $u_{0,0}$ can
be excluded from the beginning, as we explained. First order harmonics $u_{1,m}$,
describe a shift of the vesicle without changing its shape, and therefore do not play any
important role in the vesicle dynamics. The most essential role is played by the second
angular harmonic which determines mainly the vesicle shape (at small $\Delta$). As to
higher harmonics, they relax fast in comparison with the relatively slow dynamics of the
second harmonic. Therefore, the higher harmonics also do not play an essential role in
the vesicle dynamics. To avoid a misunderstanding, let us stress that the last assertion
is valid for stationary external flows. As it was discovered experimentally \cite{07KSS}
and explained theoretically \cite{07TV}, at some conditions (abrupt inversion of the
external purely elongational flow) high angular harmonics are generated, the phenomenon
is called wrinkling.

The vesicle shape depends on the strength of the external flow. In weak flows it is close
to an equilibrium one whereas in strong flows it is determined by the velocity gradient
matrix (\ref{graddec}). In this section we consider the first case. We discuss the
equilibrium vesicle shape and then develop phenomenology for the vesicle dynamics in weak
flows.

\subsection{Equilibrium}
\label{subsec:equilibrium}

In the absence of the external flow, an equilibrium shape of the vesicle can be found by
minimization of an effective free energy
 \begin{equation}
 {\cal F} = {\cal F}^{({\mathrm b})}(u)
 +\bar\sigma\, r_0^2 \Delta(u) \,,
 \label{free_energy}
 \end{equation}
where the first term is determined by the expression (\ref{Helfrich}) and $r_0^2\Delta$
is the membrane excess area expressed in terms of the displacement $u$. The Lagrange
multiplier $\bar\sigma$, related to a fixed value of the membrane area, coincides with
the equilibrium value of the surface tension. The second Lagrange multiplier (related to
the volume ${\cal V}$) is absent in Eq. (\ref{free_energy}) since we imply that the zero
angular harmonic in an expansion of the displacement $u$ is expressed via other ones from
the relation (\ref{constraint}). Therefore, the volume conservation is automatically
satisfied in our scheme.

If $\Delta$ is small, the principal contributions to the energy (\ref{Helfrich}) as well
as to the excess area are of the second order in $u$. It is convenient to write the
contributions in terms of the coefficients $u_{l,m}$ of the expansion (\ref{harmonics})
of $u(\theta,\varphi)$ over the angular harmonics:
 \begin{eqnarray}
 {\cal F}^{(2)}
 =\displaystyle\frac{\kappa}{2}
 \displaystyle\sum\limits_{l\geq2,m}
 (l+2)(l+1)l(l-1)\left|u_{l,m}\right|^2
 \nonumber \\
 +\frac{1}{2}\bar\sigma r_0^2
 \sum_{l\geq2,m}(l+2)(l-1)
 \left|u_{l,m}\right|^2\,.
 \label{secondf} \end{eqnarray}
Note, that the first angular harmonic (with $l=1$) is absent in the expansions. The
reason is that it corresponds to a vesicle shift as a whole, which does not change the
energy and the area of the vesicle. As it follows from Eq. (\ref{secondf}), the free
energy is minimal if only the second angular harmonic is excited. In this case
$\bar\sigma_0 =-6\kappa/{r}_0^2$, which is an equilibrium value of the surface tension.

Note that the second order term (\ref{secondf}) is degenerate in $m$. Therefore, in order
to determine the vesicle equilibrium shape, one should take into account terms of higher
order in the expansion of the effective free energy (\ref{free_energy}), which violate
the degeneracy. In the main approximation, it is possible to keep third order in $u$
terms, and a contribution to $u$ related to the second angular harmonics.

For a subsequent analysis, it is convenient for us to use the following real basis
  \begin{eqnarray} &&
    \psi_1 = \frac{\sqrt{5}}{4\sqrt{\pi}}(3\cos^2\theta-1) \,, \quad
    \psi_2= \frac{\sqrt{15}}{2\sqrt{\pi}}\sin(2\theta)\cos\varphi \,,
 \nonumber \\ &&
    \psi_3= \frac{\sqrt{15}}{2\sqrt{\pi}}\sin(2\theta)\sin\varphi \,, \quad
    \psi_4= \frac{\sqrt{15}}{4\sqrt{\pi}}\sin^2\theta\cos(2\varphi) \,,
 \nonumber \\ &&
    \psi_5= \frac{\sqrt{15}}{4\sqrt{\pi}}\sin^2\theta\sin(2\varphi) \,,
 \label{basispsi}
 \end{eqnarray}
instead of the traditional angular functions ${\cal Y}_{2,m}$. The functions $\psi_\mu$
are normalized as
 \begin{equation}
 \int {d}\varphi\, d\theta\, \sin\theta\,
 \psi_\mu\psi_\nu = \delta_{\mu\nu}.
 \label{psinorma}
 \end{equation}
In terms of the functions (\ref{basispsi}), the contribution to $u$ related to the second
angular harmonic, can be rewritten as follows
  \begin{equation}
    u(\theta,\varphi)    =
    \sum\limits_{\nu=1}^5u_\nu\,\psi_\nu(\theta,\varphi) \,,
  \label{psiexp}
  \end{equation}
where $u_\nu$ are some real coefficients.

Expanding the bending energy (\ref{Helfrich}) and the excess area $\Delta$ upto the third
order in $u$ and substituting there the expansion (\ref{psiexp}) one obtains
 \begin{eqnarray}
 {\cal F}^{(3)} = 12\kappa\left(
 {u_\mu u_\mu}-\Xi_{\mu\nu\lambda}{u_\mu u_\nu u_\lambda}\right)
 +\bar\sigma r_0^2 \Delta^{(3)} \,,
 \label{third} \\
 \Delta^{(3)}=2{u_\mu u_\mu}
 - 2\Xi_{\mu\nu\lambda}{u_\mu u_\nu u_\lambda}/3 \,,
 \label{dethird}
 \end{eqnarray}
where summation over repeated indices is implied and we designated
  \begin{equation}
    \Xi_{\mu\nu\lambda} = \int
    d\varphi\, d\theta\, \sin\theta\,
    \psi_\mu\psi_\nu\psi_\lambda \,.
  \label{Xi}
  \end{equation}
Components of the object $\Xi_{\mu\nu\lambda}$ are of order unity, they can be found from
the definition (\ref{Xi}) after substituting the expressions (\ref{basispsi}).

Minimizing the free energy (\ref{third}) over $u_\mu$ and determining the Lagrangian
multiplier $\bar\sigma$ from the condition $\Delta=\Delta^{(3)}$, one obtains
 \begin{equation}
 \label{exactst}
 1+\frac{\bar\sigma r_0^2}{6\kappa}=
 \frac{\sqrt{15}}{14\sqrt{\pi}} \sqrt{\Delta} \,.
 \end{equation}
This is the correction related to the third order term in the expansion of the free
energy. The minimum of the energy corresponds to a prolate uniaxial ellipsoid. If the
principal axis of the ellipsoid is directed along the $Z$-axis its shape is determined by
the expression $u_1=-\sqrt{\Delta/2}$, that is
  \begin{equation}
  u =\frac{\sqrt{5\Delta}}{4\sqrt{2\pi}}(1-3\cos^2\theta) \,.
  \label{equishape}
  \end{equation}

Substituting the expression (\ref{exactst}) into Eq. (\ref{third}) we find that the
coefficient before $u_{\mu}u_{\mu}$ term in the effective free energy is estimated as
$\kappa\sqrt\Delta$. It contains an extra small factor $\sqrt\Delta$ in comparison with
the natural estimation $\kappa$. Thus, both, second order and third order, terms in the
free energy (\ref{third}) are of the same order. That gives a formal justification of
taking the third order term into account despite the smallness of $u$.

\subsection{Weak external flow, phenomenology}
\label{subsec:weakef}

Here, we analyze the case of weak external flows that cannot significantly distort the
vesicle equilibrium shape. As we established in the preceding subsection, the equilibrium
shape of a nearly spherical vesicle is the prolate ellipsoid possessing the uniaxial
symmetry. An orientation of such ellipsoid in space can be characterized by a unit vector
$\bm n$ directed along the principal axis of the ellipsoid. If $\bm n$ is directed along
the $Z$-axis then the vesicle shape is determined by the expression (\ref{equishape}).
Note that the vectors ${\bm n}$ and $-{\bm n}$ describe the same physical state since the
ellipsoid is invariant under inversion.

One can formulate a phenomenological equation for the dynamics of ${\bm n}$ in a weak
external flow:  $\partial_t n_i=D_{ijk}\partial_k V_j$, where $\partial_k V_j$ is the
velocity gradient matrix of the external flow and $D_{ijk}$ is some tensor related to the
vesicle orientation. Due to the symmetry $\bm n\to-\bm n$ the tensor $D_{ijk}$ contains
only odd powers of ${\bm n}$. Using the relation $\bm n^2=1$, that is $n_iD_{ijk}=0$, we
arrive to the following equation
 \begin{eqnarray}
    \partial_t n_i=
    \bigl[(n_k\delta_{ij}
    -n_j\delta_{ik})/{2} \hspace{2cm}
    \nonumber \\
    +D\left(n_k\delta_{ij}/2+n_j\delta_{ik}/2
    -n_in_jn_k\right)
    \bigr]\partial_k V_j \,,
 \label{dot_n}
 \end{eqnarray}
containing a single dimensionless parameter $D$. Deriving this equation, we exploited the
fact, that the vesicle dynamics should be purely rotational  $\partial_t{\bm n}=
{\bm\omega}\times {\bm n}$ in a case of an external flow $\partial_j V_i =
-\epsilon_{ijk}\omega_k$ corresponding to a solid rotation. The factor $D$ in Eq.
(\ref{dot_n}) is dependent on relative significance of the viscous mechanisms and on the
excess area parameter $\Delta$, the dependence will be established further, see Section
\ref{sec:limit}.

\begin{figure}
 \centerline{
 \includegraphics[width=0.4\textwidth]{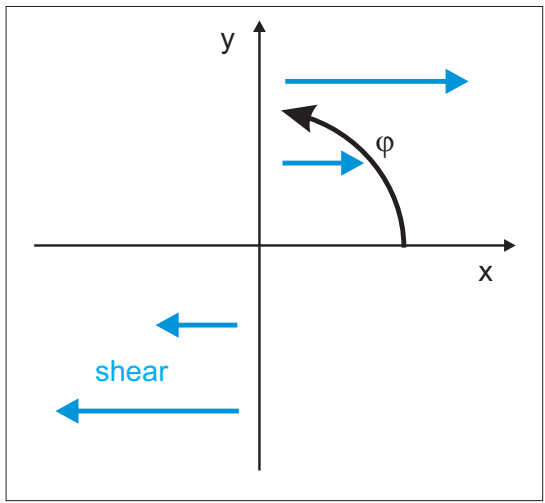} }
 \caption{Reference frame related to shear flow.}
 \label{figure:shear_flow}
 \end{figure}

For an external shear flow, it is convenient to utilize the following parametrization of
the unit vector $\bm n$
 \begin{equation}
 \label{director}
 \bm n =\left(\cos\vartheta\cos\phi,
 \cos\vartheta\sin\phi,\sin\vartheta\right)\,.
 \end{equation}
The components here are written in the Cartesian reference frame attached to the flow:
the $X$-axis is directed along the velocity and the $Z$-axis is directed opposite to the
angular velocity vector $\bm\omega$ (see Fig. \ref{figure:shear_flow} for clarification).
Substituting the expression (\ref{director}) into Eq. (\ref{dot_n}) and taking into
account that the shear velocity gradient matrix has the only non-zero component
$\partial_y V_x=\dot\gamma$ one obtains
 \begin{eqnarray} &&
 \dot\gamma^{-1}\partial_t \phi
 = (D/2)\cos(2\phi) -{1}/{2} \,,
 \label{T-TT} \\ &&
 \dot\gamma^{-1}\partial_t \vartheta =
 -({D}/{4})\sin(2\vartheta)\sin(2\phi) \,.
 \label{eqvarth}
 \end{eqnarray}
Note that the dynamics of the angle $\phi$ is separated. The equations
(\ref{T-TT},\ref{eqvarth}) resemble equations for a single polymer dynamics examined in
Ref. \cite{05CKLT}.

The equations (\ref{T-TT},\ref{eqvarth}) lead to either tank-treading or tumbling regimes
of vesicle motion. For $|D|>1$ the tank-treading regime is realized, with a steady tilt
angle (between the vector $\bm n$ and the velocity direction)
 \begin{equation}
 \phi_\ast = ({1}/{2})
 \arccos\left({1}/{D}\right) \,.
 \label{varphi_ast}
 \end{equation}
Otherwise, for $|D|<1$, the tumbling regime takes place: the vector $\bm n$ experiences a
time-periodic motion with an average rotation in the shear plane. Thus, $D=1$ corresponds
to the tank-treading to tumbling transition. As follows from Eq. (\ref{T-TT}), the
transition is realized via the saddle-node bifurcation.

\section{Perturbation expansion: general dynamics}
\label{sec:perturb}

Here, we start to carry out the program (formulated in Subsection \ref{subsec:parametr})
leading to a dynamic equation for the dimensionless displacement $u$. The program can be
realized for nearly spherical vesicles by using a generalization of the Lamb scheme. In
accordance with Lamb \cite{32Lamb} (see also Ref. \cite{65HB}), a solution of the
stationary Stokes equation can be explicitly expressed via the velocity field taken at a
sphere both for the internal and external problems. The Lamb scheme can be directly
applied to a spherical solid body immersed into a fluid or to a spherical cavity filled
up by a fluid. Then the velocity field is expressed in terms of its surface value. For a
nearly spherical vesicle the scheme has to be slightly modified. Namely, one should
express the velocity field via its value on the sphere of the radius $r_0$. The values
can be obtained by an analytical continuation of the internal and of the external
velocity fields and are slightly different for the internal and for the external
problems. The next idea is to represent the boundary conditions as an expansion over the
displacement $u$ which is small parameter for nearly spherical vesicles.

In the zeroth approximation, one can ascribe the membrane velocity directly to the sphere
$r=r_0$ ignoring deviations of the vesicle shape from the sphere. Keeping then the lowest
in $u$ terms in all expressions one passes to an equation for the displacement $u$
equivalent to the one discussed in Refs. \cite{M06,VG07}. However, as we demonstrated
in Ref. \cite{LTV07}, such approximation is not self-consistent. The problem is that it
leads to a dynamics sensitive to initial conditions. And one can overcome this
sensitivity only by accounting for high order terms in $u$.

\subsection{Closed equation}

Here, we derive an equation for the displacement $u$ in the approximation where the
membrane velocity and the boundary conditions (\ref{normal},\ref{tangent}) are related to
the sphere $r=r_0$. Corrections to the equations associated with the deviations of the
vesicle shape from the sphere are small in $u$. However, for the reasons formulated
above, we keep the leading non-linear in $u$ term in the expression for the boundary
force (\ref{surforce}).

Note, first of all, that the variational derivative of the effective free energy
(\ref{free_energy}) can be represented as
 \begin{eqnarray}
 \frac{\delta{\cal F}}{\delta u}\equiv
 r_0 \left\{-\kappa[H(H^2/2-2K) +\Delta^\perp H]
 +\bar\sigma H\right\}.
 \nonumber
 \end{eqnarray}
Therefore the boundary condition (\ref{normal}) can be rewritten as
 \begin{equation}
 \delta{\cal F}/\delta u=
 -2\tilde\sigma+ r_0 P_{in}- r_0 P_{out}.
 \label{normal1}
 \end{equation}
Here, we divided the surface tension, $\sigma$, into a homogeneous, $\bar\sigma$, and an
inhomogeneous, $\tilde\sigma$, parts. By definition, the zero angular harmonic is absent
in $\tilde\sigma$. Next, for the sphere $r=r_0$ the average curvature is $H=2/r_0$ and
$\partial^\perp_i l_n\partial^\perp_n v_i\propto \delta^\perp_{in}\partial^\perp_n
v_i=0$, that explains validity of the expression (\ref{normal1}).

To find the inhomogeneous part of the surface tension, $\tilde\sigma$, one has to use the
second boundary condition, (\ref{tangent}). Taking the derivative $\partial^\perp_i$ of
the relation (\ref{tangent}) and relating a result to the sphere $r=r_0$ one obtains
 \begin{eqnarray}
 l(l+1) \sigma_l -2\zeta(l+2)(l-1) v_{r,l}
 \nonumber \\
 =\tilde\eta\left[(l+2)(l-1)v_{r,l}+r_0^2\partial_r^2v_{r,l}\right]_{in}
 \nonumber \\
 -\eta\left[(l+2)(l-1)v_{r,l}+r_0^2\partial_r^2v_{r,l}\right]_{out},
 \label{tangent1}
 \end{eqnarray}
where $\sigma_l$ and $v_{r,l}$ are contributions to the surface tension and to the radial
velocity, respectively, associated with the $l$-th order angular harmonic. As above, the
subscripts $in$ and $out$ are related to the interior and exterior regions of the
vesicle.

Applying the Lamb scheme to the sphere $r=r_0$, one finds for the internal problem
 \begin{eqnarray}
 P_{in}=-\tilde\eta \sum_l\frac{r^l}{r_0^{l}}
 \frac{(l-1)(2l+3)}{l} \partial_t u_l,
 \label{pint1} \\
 v_r=\sum_l\left(\frac{r^{l-1}}{r_0^{l-2}}\frac{l+1}{2}
 -\frac{r^{l+1}}{r_0^{l}}\frac{l-1}{2} \right)
 \partial_t u_l.
 \label{vint1}
 \end{eqnarray}
Here, we used the condition $\partial_rv_{r}(r_0)=0$, following from the relation $l_i
l_k\partial_i v_k=0$ where $l_i=(\sin\theta \cos\varphi,\sin\theta
\sin\varphi,\cos\theta)$ is the unit vector perpendicular to the sphere $r=r_0$. We
substituted also $v_{r}(r_0)=r_0\partial_t u$, which is the kinematic relation
(\ref{kinema}) taken in the main approximation in $u$.

For the external problem, one should separate the external flow velocity $\bm V$ since it
does not tend to zero as $r\to\infty$. Writing $\bm v=\bm V+\bm w$, one obtains from the
Lamb scheme
 \begin{eqnarray}
 P_{out}=\eta \sum_l\frac{r_0^{l+1}}{r^{l+1}}
 \frac{(l+2)(2l-1)}{l+1} \partial_t u_l,
 \label{pext1} \\
 w_r=\sum_l\left(\frac{r_0^{l+1}}{r^{l}}\frac{l+2}{2}
 -\frac{r_0^{l+3}}{r^{l+2}}\frac{l}{2} \right) \partial_t u_l,
 \label{wext1}
 \end{eqnarray}
where, again, we used the incompressibility condition $\partial_rw_{r}(r_0)=0$ and the
kinematic relation $w_r=r_0\partial_t u$. However, the expressions
(\ref{pext1},\ref{wext1}) should be modified for $l=2$ because of the external flow. The
modified incompressibility condition is $\partial_r w_{r,2}(r_0)+s_{ik} l_i l_k=0$ and
the modified kinematic condition is $\partial_t u_2=w_{r,2}+r_0 s_{ik} l_i l_k$. The
modified conditions lead to the expressions
 \begin{eqnarray}
 P_{out,2}=\eta(4\partial_t u_{(2)}-5s_{ik} l_il_k){r_0^3}/{r^3},
 \label{2extp1} \\
 w_{r,2}=\left(2\partial_t u_{(2)}
 -{5}s_{ik} l_il_k/2\right){r_0^3}/{r^2}
 \nonumber \\
 -\left(\partial_t u_{(2)}-{3}s_{ik}
 l_il_k/2\right){r_0^5}/{r^4}.
 \label{2extw1}
 \end{eqnarray}

Collecting together the relations (\ref{normal1}-\ref{2extw1}) one finds a closed
equation for the displacement $u$
 \begin{equation}
 \hat a(\partial_t-\omega\partial_\varphi)u =
 10 s_{ij}l_i l_j -\displaystyle\frac{1}{\eta r_0}
 \displaystyle\frac{\delta {\cal F}}{\delta u} \,.
 \label{equ1}
 \end{equation}
where $\hat a$ is a dimensionless operator with angular components
 \begin{eqnarray}
 a_l=\frac{2l^3+3l^2+4}{l(l+1)}
 +\frac{2l^3+3l^2-5}{l(l+1)}
 \frac{\tilde\eta}{\eta}
 +\frac{l^2+l-2}{l(l+1)}\frac{4\zeta}{\eta r_0}.
 \nonumber
 \end{eqnarray}
We included into Eq. (\ref{equ1}) a dependence on the rotational part of the external
flow, which can be established by an account of the non-linear term in the kinematic
relation (\ref{kinema}). The $Z$-axis of our reference frame is implied to be directed
opposite to the angular velocity $\bm\omega$ of the external flow.

Recall that the quantity $\bar\sigma$ entering the dynamic equation (\ref{equ1}) through
Eq. (\ref{free_energy}) is the surface tension $\sigma$ averaged over angles. As
previously, $\bar\sigma$ is an auxiliary quantity ensuring the surface conservation law.
Let us stress that $\bar\sigma$ is a function of time adjusting to the current vesicle
shape. Note that the strain and the rotation parts of the external flow are separated:
the angular velocity $\bm\omega$ extends the time derivative (its effect is equivalent to
passing to the rotating reference frame) whereas the strain matrix enters the term
$s_{ij}l_i l_j$ playing a role similar to the free energy derivative. The reason is that
the elongational part of the flow leads to some viscous dissipation whereas the solid
rotation does not imply any dissipation.

One can further elaborate the equation (\ref{equ1}). First of all, one can keep in the
effective free energy (\ref{free_energy}) terms of the second and of the third order in
$u$. Higher order terms in ${\cal F}$ are small since $u\ll1$. The reason why the third
order term should be kept besides the second order one is explained above. Next, the
expansion of the term $s_{ij}l_i l_j$  does not contain any angular harmonics with $l>2$,
so this term does not push $u$ outside the $l=2$ subspace. Therefore the higher order
angular harmonics in $u$ die out after a finite time due to the last term in
(\ref{equ1}). Therefore one can use a reduced equation where $u$ contains only second
order angular harmonics. The operator $\hat a$ in this case is reduced to a constant
 \begin{equation}
 a=\frac{16}{3}
 \left(1+\frac{23}{32}\frac{\tilde\eta}{\eta}+
 \frac{\zeta}{2\eta r_0}\right) \,,
 \label{defa}
 \end{equation}
depending on the viscous ratios. The constant $a$ can be called a generalized viscosity
contrast. Note that the limit $a\to\infty$ (where the viscosity of the internal fluid or
the membrane viscosity tend to infinity) should correspond to a solid body behavior of
the vesicle.

One can substitute the expression (\ref{third}) into the equation (\ref{equ1}) and then
project the equation on the subspace $l=2$. Then the equation (\ref{equ1}) is reduced to
 \begin{eqnarray}
 a\left[\partial_t u_\mu-\omega(\partial_\varphi u)_\mu\right] =
 10 (s_{ij}l_i l_j)_\mu \hspace{2cm}
 \label{main} \\
 -\frac{24}{\eta{r}_0}\left[
 \left(\frac{\kappa}{r_0^2}+\frac{\bar\sigma}{6}\right)u_\mu
 -\left(\frac{3\kappa}{2 r_0^2}+\frac{\bar\sigma}{12}\right)
 \Xi_{\mu\nu\lambda}u_\nu u_\lambda \right],
 \nonumber
 \end{eqnarray}
where the subscripts in $(\partial_\varphi u)_\mu$ and $(s_{ij}l_i l_j)_\mu$ designate
projections to the basis (\ref{basispsi}) calculated in accordance with Eq.
(\ref{psinorma}).

\subsection{Rescaled equation}

The factor $\bar\sigma$ in Eq. (\ref{main}) should be extracted from the condition $2
u_\mu u_\mu =\Delta$, which is the area conservation law written in the main
approximation in $\Delta$. Then one obtains
 \begin{eqnarray}
 a\left[\partial_t u_\mu-\omega(\partial_\varphi u)_\mu\right]
 =\left(\delta_{\mu\rho}
 -\frac{2u_\mu u_\rho}{\Delta}\right)
 \nonumber \\
 \times\left[10 (s_{ij}l_i l_j)_\rho
 +\frac{24\kappa}{\eta{r}_0^3}\,
 \Xi_{\rho\nu\lambda}u_\nu u_\lambda \right],
 \label{main1}
 \end{eqnarray}
where terms of the order $s u$ are neglected. In accordance with Eq. (\ref{decompos}),
the right-hand side of Eq. (\ref{main1}) is a sum of two terms, proportional to $s_{ij}$
and to $\kappa$. Note that the term proportional to $\kappa$, is of the second order in
$u$ (the first order term is absent), that justifies keeping this high-order term in the
expansion of the free energy.

Deriving the equation (\ref{main1}) we neglected terms of order $su$. They are much less
compared to the term with $\kappa$ provided $s\ll \kappa\sqrt\Delta/(\eta r_0^3)$. This
is the applicability condition of the equation. However, below we demonstrate that
results obtained from the equation (\ref{main1}) can be extended to stronger flows.

After some rescaling the equation (\ref{main1}) can be rewritten as
 \begin{eqnarray}
 \label{diless}
 \left(\!\tau_\ast\,\partial_t
 \!-\! \frac{S\Lambda}{2}\partial_\varphi\!\right)\! U_\mu
 \!=\!(\delta_{\mu\rho}\!-\!U_\mu U_\rho)
 (S_\rho\!+\!\tilde\Xi_{\rho\nu\lambda}U_\nu U_\lambda),
   \\
 U_\mu ={\sqrt{2}u_\mu}/{\sqrt\Delta}, \quad
 U_1^2+ \dots +U_5^2=1,
 \label{Unorma}
 \end{eqnarray}
where $\tilde\Xi = ({7\sqrt{\pi}}/{\sqrt{5}})\Xi$. The parameters in the equation
(\ref{diless}) are defined as follows
 \begin{eqnarray}
  \label{tau_ast}
 \tau_\ast
 =\frac{7\sqrt{\pi}}{12\sqrt{10}}
 \frac{a \eta r_0^3}{\kappa \sqrt{\Delta}},
 \\
 \label{selfsim}
    S  = \frac{14\pi}{3\sqrt{3}}\frac{s \eta r_0^3}{\kappa\Delta}, \qquad
    \Lambda = \frac{\sqrt{3}}{4\sqrt{10\pi}}\frac{\sqrt{\Delta}a\omega}{s},
 \end{eqnarray}
where $s^2=s_{ij}s_{ij}/2$.

The ``vector'' $S_\mu$ in Eq. (\ref{diless}) has an absolute value $S$, given by Eq.
(\ref{selfsim}), its ``direction'' is determined by the projections of the object
$s_{ij}l_i l_j$ to the basis (\ref{basispsi}) that is the ``direction'' is determined by
the structure of the strain matrix $s_{ij}$. Therefore the two parameters, $S$ and
$\Lambda$, together with the ``direction'' of the ``vector'' $S_\mu$ completely determine
a character of the vesicle dynamics in the external flow.

The quantity $\tau_\ast$ is the characteristic time scale of the vesicle relaxation. In
comparison with the combination $\eta r_0^3/\kappa$, related to the external fluid, the
time $\tau_\ast$ contains an additional factor $a$, reflecting the contributions of the
internal fluid viscosity and the membrane viscosity into the relaxation, and also the
factor $\Delta^{-1/2}$. This extra factor reflects the slowness of the second order
angular harmonic relaxation related to the degeneracy of the second order free energy in
$m$. Due to the degeneracy the relaxation is determined by the third order term in $u$ in
the effective free energy, which contains a smallness $\Delta^{1/2}$ in comparison with
the energy of higher angular harmonics. Therefore the adiabaticity condition, enabling
one to use the stationary Stokes equation, can be written as $\tau_\ast\gg \varrho
r_0^2/\eta$ that is $a\eta^2 r_0\gg \rho\kappa\sqrt\Delta$. The inequality is valid
because of the large value of the radius $r_0$ (in comparison with the molecular length)
and  the small value of $\Delta$.

The parameter $S$ characterizes the relative strength of the external flow. The
expression (\ref{selfsim}) for $S$ can be explained as follows. The external viscous
surface force $\eta s$ should be balanced by the surface tension $\bar\sigma$ times a
variation of the vesicle curvature which is estimated as $\sqrt\Delta/r_0$. Therefore
$\bar\sigma\sim \eta s r_0/\sqrt\Delta$. As we have established in Subsection
\ref{subsec:equilibrium}, see Eq. (\ref{exactst}), the characteristic equilibrium surface
tension can be estimated $\bar\sigma_0\sim \kappa\sqrt\Delta /r_0^2$. Ratio of the
surface tension values is therefore estimated as $S$: $S\sim\bar\sigma/\bar\sigma_0$. The
applicability condition $s\ll \kappa\sqrt\Delta/ (\eta r_0^3)$ of the equations
(\ref{main1},\ref{diless}) can be rewritten as $S\ll 1/\sqrt\Delta$, in terms of $S$.

The parameter $\Lambda$ determines the relative strength of the rotational part of the
external flow. Note, that $\omega\tau_\ast \sim S \Lambda$. The condition $\Lambda\sim1$
corresponds to the angular velocity whose effect is comparable with the effect of the
strain, the condition gives $\omega\sim s/(a\sqrt\Delta)$. This characteristic angular
velocity does not coincide with the characteristic value of $s$, that stresses again
different roles of the rotational and of the elongational parts of the external flow.

\subsection{Very strong external flows}
\label{subsec:very}

Let us consider the case of very strong external flows, $S\gg 1/\sqrt\Delta$, that is
$s\gg \kappa\sqrt\Delta/(\eta r_0^3)$. As we already explained, the membrane surface
tension, which is determined by the balance between the viscous surface force $\eta s$
and the product of surface tension $\bar\sigma$  and the variation of the vesicle
curvature $\sqrt\Delta/r_0$, is estimated as $\bar\sigma\sim \eta s r_0/\sqrt\Delta$.
Therefore for the external flows, we consider here, $\bar\sigma\gg \kappa/r_0^2$. The
inequality implies that the leading role in the vesicle reaction to the external flow is
played by the surface tension.

Based on the inequality $\bar\sigma\gg \kappa/r_0^2$, one could try to neglect the terms
with the module $\kappa$ in Eq. (\ref{main}) to obtain an equation independent of the
curvature energy (\ref{Helfrich}). That would correspond to omitting the contribution
$\bm v^{(\kappa)}$ in the decomposition (\ref{decompos}). However, the resulting equation
is incorrect. First, there are additional terms in the equation for $u$, originating from
the regular expansion over $u$ and related to an account of deviations of the vesicle
shape from the sphere $r=r_0$. The additional terms, which can be estimated as $su$, are
of the same order as kept in Eq. (\ref{main}). We ignored the terms in the equation
(\ref{main1}) for $u$ at $s\ll\kappa \sqrt\Delta/ (\eta r_0^3)$ since then the term $\sim
su$ is small. Now the additional terms should be taken into account. Second, as we will
see in the case of the planar flow, the truncated equation (without $\kappa$-terms)
cannot be used for describing the vesicle dynamics because of its internal properties.

If the terms with the module $\kappa$ are neglected in the boundary conditions
(\ref{normal},\ref{tangent}) then they appear to be invariant under the transformation
$\bm v\to-\bm v,\ P\to-P,\ \sigma\to-\sigma$. The stationary Stokes equation
$\eta\nabla^2 \bm v=\nabla P$ and the boundary condition (\ref{sincomp}) are also
invariant under the transformation. The kinematic relation (\ref{kinema}) becomes
invariant under the transformation if one adds the rule $t\to-t$. Therefore, in this
approximation the backward in time evolution of the displacement $u$ is equivalent to the
direct evolution in the external flow with the velocity $-\bm V$. Since the vesicle
dynamics is determined by the matrix (\ref{graddec}), the transformation $\bm V\to-\bm V$
is equivalent to space inversion. That produces some additional symmetry leading to
essential consequences for planar velocity fields.

\section{Planar velocity field}
\label{sec:planar}

In this section we discuss in more details the case of a planar external velocity field,
when the velocity vector $\bm V$ lies in a plane and is independent of a coordinate
normal to the plane. Then the velocity gradient matrix $\partial_j V_i$ has nonzero
elements only in the plane being a $2d$ matrix. We choose $X$ and $Y$ axes of our
reference system parallel to the plane and assume without the loss of generality that the
diagonal elements of the matrix $\partial_j V_i$ are zero. Two non-diagonal components of
the matrix $\partial_j V_i$ completely determine the flow, they can be parameterized in
terms of the strain $s$ and the angular velocity $\omega$ as $\partial_y V_x = s+\omega$
and $\partial_x V_y = s-\omega$. In particular, for an external shear flow the only
nonzero element of the matrix is $\partial_y V_x = \dot\gamma$ and $\omega = s =
\dot\gamma/2$.

For a planar velocity field, the only nonzero element of $S_\mu$ in Eq. (\ref{diless}) is
$S_5 = S$. Thus only two parameters in Eq. (\ref{diless}), $S$ and $\Lambda$, completely
determine the type of the vesicle dynamics. One of the main goals of this paper is to
construct a ``phase diagram'' in the plane $S-\Lambda$, which indicates a type of the
vesicle motion for a given pair of these parameters.

Very strong flows, characterized by $S\gg 1/\sqrt\Delta$ need a special consideration due
to the reasons formulated above. It is presented in a separate subsection. Surprisingly,
the case can be analyzed in the framework of the same equations
(\ref{main1},\ref{diless}).

\subsection{General consideration}
\label{subsec:general}

For the planar flow, it is convenient to pass from the variables $U_\mu$ (\ref{Unorma})
to another set of variables, ``angles'' $\Phi,\Theta,\Psi,{J}$, defined as
 \begin{eqnarray}&
 \label{U_representation}
    U_1 = \sin\Theta\cos {J},
    \\&
    U_2 = \sin\Theta\,\sin {J}\,\cos\Psi,
    \quad\nonumber
    U_3 = \sin\Theta\,\sin {J}\,\sin\Psi,
    \\&
    U_4 = \cos\Theta\cos(2\Phi),
    \quad\nonumber
    U_5 = \cos\Theta\sin(2\Phi),
 \end{eqnarray}
where $\Theta$ varies from $-\pi/2$ to $\pi/2$ and $J$ varies from $0$ to $\pi/2$. The
representation (\ref{U_representation}) satisfies the normalization condition
(\ref{Unorma}) and, correspondingly, consists of four variables instead of five
components $U_\mu$.

As follows from the equations (\ref{diless}), there is a solution with ${J}=0$. Actually,
it is a consequence of general symmetry $\theta\to-\theta$, which is present in the
vesicle equations in the planar flow. Due to this symmetry, there exists a symmetric in
$\theta$ solution of $u$, that corresponds to $J=0$ as it follows from Eqs.
(\ref{basispsi},\ref{psiexp},\ref{U_representation}). A special analysis is required to
analyze the stability of the solution. Our numerical simulations demonstrated the
stability. That is why we mainly limit ourselves to the investigation of this case. At
$J=0$ the displacement $u$ is
 \begin{equation}
 u\propto \cos\Theta \sin^2\theta \cos(2\varphi-2\Phi)
 +\sin\Theta \frac{1-3 \cos^2\theta}{\sqrt3} .
 \label{uthph}
 \end{equation}
Therefore the ``angle'' $\Phi$ characterizes the vesicle orientation in the $X-Y$ plane
whereas the ``angle'' $\Theta$ determines the vesicle shape. In the case $J=0$ the system
of equations (\ref{diless}) is reduced to the following two equations
 \begin{eqnarray}
 \tau_\ast\partial_t\Theta = -S\sin\Theta\sin(2\Phi) +\cos(3\Theta),
 \label{eq_on_Theta} \\
 \label{eq_on_Phi}
 \tau_\ast\partial_t\Phi =
 \frac{S}{2} \left[\frac{\cos(2\Phi)}{\cos\Theta}
 - \Lambda \right].
 \end{eqnarray}
Note that the last, nonlinear in $u$, summand in (\ref{main1}) produces the term
$\cos(3\Theta)$ in Eq. (\ref{eq_on_Theta}) and no term in Eq. (\ref{eq_on_Phi}).


 \begin{figure}
 \centerline{
 \includegraphics[width=0.55\textwidth]{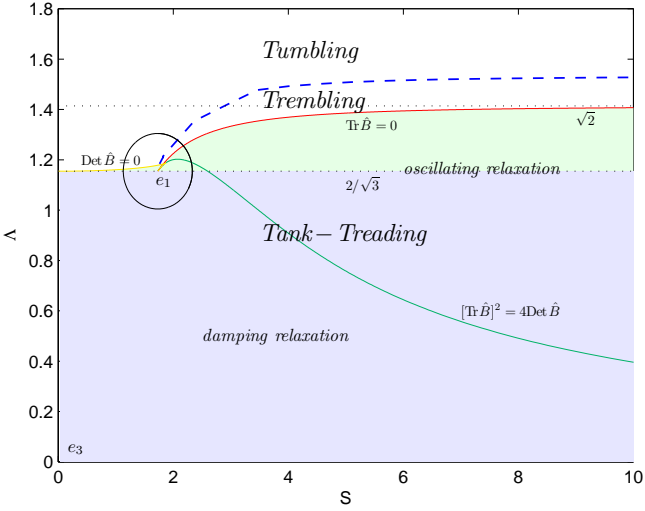} }
 \caption{Phase diagram}
 \label{figure:phase_diagram}
 \end{figure}
 \begin{figure}
 \centerline{
 \includegraphics[width=0.55\textwidth]{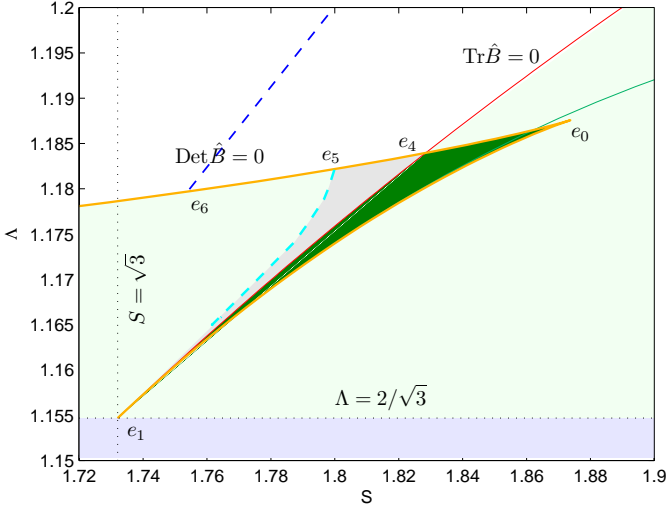} }
 \caption{Phase diagram in a vicinity of the special point.}
 \label{figure:Triangle}
 \end{figure}

Let us find a region of parameters $S$ and $\Lambda$ where the equations
(\ref{eq_on_Theta},\ref{eq_on_Phi}) admit stable stationary points, the case corresponds
to the tank-treading vesicle motion. Equating to zero the right hand sides of the
equations, one finds relations determining a stationary point (for given parameters $S$
and $\Lambda$). In order to investigate its stability one should linearize the equations
(\ref{eq_on_Theta},\ref{eq_on_Phi}) near the stationary point to obtain
 \begin{equation}
 \tau_\ast\partial_t(\delta\Theta,\delta\Phi)
 =\hat B(\delta\Theta,\delta\Phi).
 \label{stability}
 \end{equation}
The stationary point is stable, if both eigenvalues of the matrix $\hat B$ have negative
real parts. Thus the stability conditions are $\mathrm{tr}\ B<0$ and $\mathrm{det}\ B>0$.
A region in the $S$-$\Lambda$ plane where stationary stable points exist is shown on the
Fig. \ref{figure:phase_diagram}. The region below the line $\Lambda = 2/\sqrt{3}$
corresponds to positive values of $\Phi$. The tank-treading regime above the line
$\Lambda=2/\sqrt{3}$ leads to negative values of $\Phi$ for $S>\sqrt{3}$.

For the upper region in Fig. \ref{figure:phase_diagram} the attractors of the system
(\ref{eq_on_Theta},\ref{eq_on_Phi}) are limit cycles. They correspond to either tumbling
or trembling behavior. The difference is illustrated in Fig. \ref{figure:stp01} where the
$\Theta-\Phi$ atlas is plotted. The tumbling regime corresponds to a cycle separating the
atlas into two regions each containing a pole $\Theta=\pm\pi/2$, and the trembling regime
corresponds to a cycle separating the atlas into two regions one of which does not
contain any of the poles. In the tumbling regime the ``angle'' $\Phi$ grows without a
limit whereas in the trembling regime it varies in a restricted domain. A transition line
from tumbling to trembling, obtained numerically, is depicted by a dashed line in Fig.
\ref{figure:phase_diagram}.

The diagram has a complicated structure near the special point $S=\sqrt3$,
$\Lambda=2/\sqrt3$. A vicinity of the point is depicted on Fig. \ref{figure:Triangle},
where the regions of coexistence of two different stable points and of a stable point and
of a limit cycle are shown. More detailed description is given below.

To avoid a misunderstanding, note that there is an additional region in the $S-\Lambda$
plane where stationary solutions exist, which are stable in terms of the variables
$\Theta$ and $\Phi$. However, a stability investigation in the framework of the complete
equation (\ref{diless}) shows that these solutions are unstable in the extended space
with the variables ${J}$ and $\Psi$. Therefore these solutions cannot be realized as the
tank-treading motion in real systems.

\subsection{Tank-treading to trembling transition}

The tank-treading to trembling transition is determined by the condition $\mathrm{tr}\
B=0$. The corresponding curve on the $S-\Lambda$ plane starts from the above special
point $S=\sqrt3$, $\Lambda=2/\sqrt3$ (the point $e_1$ on Fig. \ref{figure:Triangle}) and
goes to the right. This curve, marked as red, is described by the equation
 \begin{equation}
 \label{hopf}
 \Lambda^\ast = \sqrt{2}\sqrt{1-1/S^2},
 \end{equation}
where $S$ varies from $\sqrt3$ to $\infty$. Above the red curve, at $\Lambda>
\Lambda^\ast$, the stationary point loses its stability via a Hopf bifurcation. Let us
establish characteristics of the bifurcation.

Expanding the equations (\ref{eq_on_Theta},\ref{eq_on_Phi}) near the point (\ref{hopf}),
one finds the equation for a complex variable $z$
 \begin{equation}
 \tau_\ast\partial_t z = \varepsilon z
 - i \sqrt{S^2-3}\, z -K|z|^2z,
 \label{hopf1}
 \end{equation}
where
 \begin{eqnarray}
 \varepsilon=\frac{8S\sqrt{S^2-1}}{S^2-3}(\Lambda-\Lambda^\ast), \
 K = \frac{2\sqrt{2}\sqrt{S^2-1}(S^2+5)}{(S^2-3)^{3/2}}.
 \nonumber \end{eqnarray}
The variable $z$ is expressed via the deviations of the ``angles'' from their stationary
values as
 \begin{eqnarray}
    \left(\!
    \begin{array}{c}
       \sqrt 2 \delta \Theta \\
        \delta \Phi
    \end{array} \!
    \right)
    \!=\!\left(
    \begin{array}{cc}
        \mu^{-1}\!\!-i\mu & \mu^{-1}\!\!+i\mu \\
        -\mu^{-1}\!\!-i\mu & -\mu^{-1}\!\!+i\mu
    \end{array}
    \right)
    \left(
    \begin{array}{c}
    z \\
    z^\ast
    \end{array}
    \right),
 \nonumber \\
 \mu=\sqrt[4]{\frac{\sqrt{S^2-1}
 +\sqrt{2}}{\sqrt{S^2-1}-\sqrt{2}}}.
 \hspace{2cm} \nonumber
 \end{eqnarray}

Above the transition line, at $\Lambda>\Lambda^\ast$, the vesicle motion is described by
a limit cycle with the radius proportional to $\sqrt{\Lambda-\Lambda^\ast}$, near the
transition curve. This motion corresponds to trembling since the radius is small, and the
corresponding limit cycle cannot surround a pole, see the atlas in Fig.
\ref{figure:stp01}.

Note the critical dependence of all parameters in Eq. (\ref{hopf1}) on $S-\sqrt3$. Taking
into account the critical dependence, one concludes that near the line the amplitudes of
the $\Theta$ and $\Phi$ variations can be estimated as $\sqrt{\Lambda-\Lambda^\ast}$,
without a critical dependence on $S-\sqrt3$.

A vicinity of the special point $e_1$ needs an additional analysis since the frequency
$\tau_\ast^{-1}\sqrt{S^2-3}$ of the Hopf bifurcation tends to zero at the point and the
approximation leading to the equation (\ref{hopf1}) is not valid there.

\subsection{Tank-treading to tumbling transition}

The transition from tank-treading to tumbling is determined by the condition
$\mathrm{det}\ B=0$. The transition curve on the $S-\Lambda$ plane, designated as orange,
has a complicated shape, it can be described in a parametric form
 \begin{eqnarray}
    S = \frac{\zeta^2 \sqrt{15\!-\!32 \zeta^2\!+\!16 \zeta^4}}{1-\zeta^2}, \
    \Lambda = \frac{\sqrt{\!-\!8 \zeta^4\!+\!12 \zeta^2\!-\!3}}
    {\zeta^2 \sqrt{5-4 \zeta^2}},
  \label{param}
 \end{eqnarray}
where the parameter $\zeta$ varies from $1/\sqrt{2}$ upto $\sqrt{3}/2$. The boundary
value $\zeta= 1/\sqrt{2}$ corresponds to the above special point $S=\sqrt3$,
$\Lambda=2/\sqrt3$, and the boundary value $\zeta=\sqrt{3}/2$ corresponds to the point
$S=0$, $\Lambda=1/\sqrt 3$. The expression for $S$ has a maximum at $\zeta_0=
\sqrt{1-2^{-4/3}}$, for $\zeta$ close to $\zeta_0$ we obtain $S \approx 1.8737 - 46.97
(\zeta-\zeta_0)^2$. The value $\zeta=\zeta_0$ corresponds to the turning point $e_0$ on
Fig. \ref{figure:Triangle}.

To be more precise the condition $\mathrm{det}\ B=0$ determines a stability boundary of
the tank-treading regime. The region of parameters $1/\sqrt{2}<\zeta<\zeta_0$,
determining the part of the orange curve going from the point $e_1$ to the point $e_0$,
corresponds to a transition from one tank-treading regime to another one. That is why the
dark green region on Fig. \ref{figure:Triangle} corresponds to coexisting two
tank-treading regimes. At passing from the region through the red curve the Hopf
bifurcation occurs. However, the bifurcation takes place for one of two possible
tank-treading regimes, the other one remains stable. Therefore there exists a region of
coexistence of the tank-treading and trembling, the region is marked as fuchsia on Fig.
\ref{figure:Triangle}. Its left boundary, corresponding to an instability of the
trembling motion, is found numerically.

Let us consider now the region of parameters $\zeta_0<\zeta<\sqrt{3}/2$, giving the upper
part of the orange curve. We are interested in the dynamics of the deviations
$\delta\Theta$, $\delta\Phi$ from the stationary values of the ``angles''. An analysis
shows that there exists a linear combination $\xi$ of the deviations possessing a slow
dynamics, that is the characteristic relaxation time of $\xi$ scales as
$\sqrt{\delta\Lambda}$ where $\delta\Lambda=\Lambda -\Lambda(S)$ and $\Lambda(S)$ is the
value of the parameter $\Lambda$ at the transition curve, determined by the expressions
(\ref{param}). Then, using adiabaticity, it is possible to formulate a closed equation
for the parameter $\xi$ which is
 \begin{equation}
 \tau_\ast\partial_t \xi=
 \delta\Lambda\,F_1 + F_2\, \xi^2,
 \label{qwr}
 \end{equation}
valid at small $\delta\Lambda$ and $\xi$. The expression (\ref{qwr}) is characteristic of
a saddle-node bifurcation.

The parameters of the saddle-node bifurcation (\ref{qwr}) have critical behavior near the
boundary points $\zeta=\zeta_0$ and $\zeta=\sqrt{3}/2$ which can be expressed as
\begin{equation}
    F_1 = f_1\sqrt{3-4\zeta^2},
    \qquad
    F_2 = f_2\sqrt{3-4\zeta^2} \left(\zeta - \zeta_0\right),
 \label{critsn}
 \end{equation}
where $f_1$ and $f_2$ are functions of $\zeta$ varying less than by $25\%$ as $\zeta$
runs from $\zeta_0$ to $\sqrt{3}/2$. In the normalization where $\xi^2=(\delta\Theta)^2
+(\delta\Phi)^2$ one finds $f_1 \approx 4.5$ and $f_2 \approx 60$ at $\zeta\to\zeta_0$.
The function $F_2$ tends to zero as $\zeta\to\zeta_0$. Therefore higher order terms in
the equation for $\xi$ should be taken into account near the point.

The limit cycle which is a result of the saddle-node bifurcation, destroying the
tank-treading regime, could be unstable in its turn. The situation is realized between
the points $e_0$ and $e_5$, see Fig. \ref{figure:Triangle}. Above the segment $e_0e_4$ a
final result of the instability is another tank-treading regime continuously continuing
to larger $S$. Above the segment $e_4e_5$ it is trembling, characterized by a limit cycle
which does not pass through the stationary point. After the point $e_5$ (to the left from
the point) the tank-treading regime is destroyed and a limit cycle  passing through the
stationary point is formed. Above the segment $e_5e_6$ the cycle corresponds to
trembling, otherwise it corresponds to tumbling.

Both functions, $F_1$ and $F_2$, tend to zero as $\zeta\to\sqrt{3}/2$. This is the limit
of weak external flows, where the vesicle relaxation rate is proportional to the strength
of the flow. Then the right-hand side of the equation should be proportional to $S$ which
behaves like $S\propto \sqrt{3-4\zeta^2}$ as a consequence of Eq. (\ref{param}). That
explains the dependence $F_1,F_2\propto \sqrt{3-4\zeta^2}$. However, the limit of weak
flows needs an additional analysis since, in accordance with results of Subsection
\ref{subsec:weakef}, there are two soft degrees of freedom corresponding to rotations of
the equilibrium uniaxial ellipsoid. We postpone the analysis to the next section.

\section{Limit cases}
\label{sec:limit}

We established a general picture of the vesicle dynamics in an external flow which
appears to be rich of different types of behavior. Particularly, the phase diagram
depicted in Fig. \ref{figure:phase_diagram} contains different domains and has a
complicated structure. The situation is simplified for different limiting cases which can
be analyzed in more detail.

For example, the equation (\ref{eq_on_Phi}) leads to the conclusion that in the case
$\Lambda\to\infty$ the vesicle rotates with the angular velocity $\omega$. For external
flows with comparable strain and rotational parts vesicle rotation is quite natural,
since in accordance with the definition (\ref{selfsim}) the limit $\Lambda\to\infty$ is
achieved either at $a\to\infty$ or at $s\to0$. The first case corresponds to a solid body
behavior of the vesicle, so one should reproduce the classical Jeffery's result
\cite{jeffery}, which predicts, that for external flows with $\omega>\sqrt{\Delta}s$ the
tumbling regime supersedes the tank-treading one. The second case corresponds to a purely
rotational external flow, where the fluid rotates as a whole with all inclusions.

Below, we analyze more complicated limit cases.

\subsection{Purely elongational flow}

The purely elongational flow is realized provided the angular velocity $\omega$ is equal
to zero, that is $\partial_y V_x = \partial_x V_y = s$. Therefore, in our designations,
the elongation is directed along the main diagonal in the $X-Y$ plane.

The condition $\omega=0$ leads to $\Lambda=0$, in accordance with the definition
(\ref{selfsim}). In this case the system of equations (\ref{eq_on_Theta},\ref{eq_on_Phi})
has a stable stationary point $\Phi_0,\Theta_0$, determined by the relations
 \begin{equation}
 \Phi_0=\pi/4,  \quad
 S \sin\Theta_0=\cos(3\Theta_0).
 \label{PTelon}
 \end{equation}
The ``angle'' $\Theta_0$ monotonically decreases from $\pi/6$ to zero as $S$ increases
from zero to infinity. The value $\Phi_0=\pi/4$ is quite natural since it corresponds to
the vesicle orientation along the elongation direction, as is seen from Eq.
(\ref{uthph}).

The stability of the point (\ref{PTelon}) can be easily established from the linearized
equations
 \begin{eqnarray}
 \tau_\ast\partial_t\delta\Phi
 =-\frac{S}{\cos\Theta_0}\delta\Phi, \hspace{2cm}
 \label{Phist} \\
 \tau_\ast\partial_t \delta\Theta=
 -\left[S\cos\Theta_0+3\sin(3\Theta_0)\right]\delta\Theta.
 \label{Thist}
 \end{eqnarray}
In the limit $S\gg1$ both ``angles'' relax to their equilibrium values with the same rate
$8\sqrt{10\pi}\, s(3\sqrt{3\Delta}\ a)^{-1}$.

Recently a wrinkling effect was observed in purely elongational flows at a sudden
invertion of the elongation direction, see Ref. \cite{07KSS}. The effect can be explained
in the framework of our scheme, the corresponding analysis is presented in Ref.
\cite{07TV}.

\subsection{Weak external flows}
\label{subsec:weak_external_flow_addition}

Let us consider weak external flows characterized by the condition $S\ll1$. We have
already discussed the case in Section \ref{subsec:weakef} from the phenomenological point
of view. We can analyze the case in terms of the ``angles'' $\Theta$ and $\Phi$ and then
establish a value of the phenomenological constant $D$, introduced in Eq. (\ref{dot_n}).

As follows from Eq. (\ref{eq_on_Theta}), for $S\ll1$ the ``angle'' $\Theta$ is close to
$\pi/6$, which is a stable point of the equation. Substituting the value $\Theta=\pi/6$
into Eq. (\ref{eq_on_Phi}) one obtains a closed equation for the ``angle'' $\Phi$
 \begin{equation}
    \tau_\ast \partial_t\Phi =(S/\sqrt{3})\cos(2\Phi) - S\Lambda/2.
 \label{weak1}
 \end{equation}
If $\Lambda<2/\sqrt3$ then the equation (\ref{weak1}) has a stationary point
 \begin{equation}
  \Phi=\frac{1}{2}\arccos \left(\frac{\sqrt 3\,\Lambda}{2}\right),
 \label{weak3}
 \end{equation}
which is stable. Otherwise $\Phi$ increases unlimited, that corresponds to the tumbling
regime. Therefore $\Lambda=2/\sqrt3$ is the transition point from tank-treading to
tumbling.

If $\Theta=\pi/6$ then the expression (\ref{uthph}) describes a prolate uniaxial
ellipsoid with the principal axis directed along the vector (\ref{director}) with
$\phi=\Phi$ and $\vartheta=0$. Comparing then the equation (\ref{weak1}) with Eq.
(\ref{T-TT}) (obtained for a shear flow with $s=\omega=\dot\gamma/2$) one finds
 \begin{equation}
 D=\frac{8\sqrt{10\pi}}{3 a\sqrt{\Delta}}.
 \label{weak2}
 \end{equation}
As it should be, the transition point $D=1$ from tank-treading to tumbling corresponds to
$\Lambda= 2/\sqrt3$. Let us stress that the value (\ref{weak2}) is independent of the
character of the external flow. Therefore the equation (\ref{dot_n}) with (\ref{weak2})
is correct for any external flow $\bm V$.

Note that in the ``solid body'' limit $a\to\infty$ (where the viscosity of the internal
fluid or the membrane viscosity tend to infinity) the quantity $D$ tends to zero. Account
of higher order terms in $\Delta$ gives, that in the solid limit $D$ stops decrease at
value order of $\sqrt{\Delta}$. Diminishing of $D$ leads to a solid rotation of the
vesicle in particular case of external shear flow as follows from Eqs.
(\ref{T-TT},\ref{eqvarth}). The behavior corresponds to the classical result of Jeffery
\cite{jeffery}, who demonstrated that a solid ellipsoid rotates in an external planar
flow, provided $\omega>\sqrt{\Delta}s$.

\subsection{Strong external flows, truncated equations}

In the case of strong external flows, where $S$ is large, the last (non-linear in $U$)
term in the right-hand side of Eq. (\ref{diless}) is small in comparison with the first
one. If to omit this last term we pass to a truncated equation. In terms of the variables
introduced by Eq. (\ref{U_representation}) the truncated equation is written as a system
of equations
 \begin{eqnarray}&&
 \nonumber
    (\tau_\ast/S)\,\partial_t\Theta
    = -\sin\Theta\sin(2\Phi),
    \\[7pt]&&
    (\tau_\ast/S)\,\partial_t \Phi
    =
    \displaystyle\frac{1}{2}
    \left[\displaystyle\frac{\cos(2\Phi)}{\cos\Theta} - \Lambda \right],
    \nonumber
    \\[7pt]&&
    (\tau_\ast/S)\,\partial_t \Psi = \Lambda/2, \qquad
    \partial_t {J}=0,
    \label{truncated}
 \end{eqnarray}
homogeneous in $S$. The system (\ref{truncated}) corresponds to the limit case considered
by Misbah \cite{M06} and Vlahovska and Gracia \cite{VG07}. In the subsection we examine
solutions of the system (\ref{truncated}). A relation to observable behavior of vesicles,
which is not straightforward, is discussed in the next subsection.

The system (\ref{truncated}) leads to conservation of two quantities, ${J}$ and an
additional integral $\Upsilon$, which can be introduced via the relation
 \begin{equation}
  \frac{\sin \Upsilon}{\Lambda-\cos \Upsilon}
  =\frac{\sin\Theta}{\Lambda- \cos\Theta\cos(2\Phi)}.
 \label{second}
 \end{equation}
For definiteness, we choose a root of the equation (\ref{second}) lying in the domain
$|\Upsilon|<\arccos(1/\Lambda)$. Existence of two integrals of motion, ${J}$ and
$\Upsilon$, implies that a character of an evolution, described by the system
(\ref{truncated}), depends on initial conditions (determining the values of the
integrals).

Thanks to existence of two integrals of motion, the system of equations (\ref{truncated})
can be completely integrated. For the purpose we introduce a variable
 \begin{equation}
 \rho = \exp\left\{S/\tau_\ast\int^t_0\mathrm{d}t'\,
 U_5(t')\right\}.
 \label{rho}
 \end{equation}
It turns out, that for arbitrary initial conditions a solution passes through a point,
where $U_5 = 0$. It is convenient to choose initial time, $t=0$, as a moment, when $U_5 =
0$ and $U_4 = \cos \Upsilon>0$. Then the initial conditions are $\rho = 1$ and
$\partial_t \rho =  0$ and one derives from the system (\ref{truncated}) the following
equation
 \begin{equation}
 \label{eq_on_rho}
    (\tau_\ast/S)^2\,\partial_t^2 \rho   =
    -(\Lambda^2-1)\rho +
    \Lambda^2 - \Lambda \cos \Upsilon ,
 \end{equation}
which can be obviously solved explicitly. The parameters $\Phi$ and $\Theta$ are
expressed via the variables $\rho$ and $\Upsilon$ as
 \begin{eqnarray}&&
    U_4 = \cos\Theta\cos(2\Phi)
    =
    \Lambda\rho - \Lambda + \cos\Upsilon
    \\ &&
    U_5 = \cos\Theta\sin(2\Phi)
    =
    (\tau_\ast/S)\partial_t \rho/\rho.
 \label{PhiandTheta}
 \end{eqnarray}

 \begin{figure}
 \centerline{
 \includegraphics[width=0.4\textwidth]{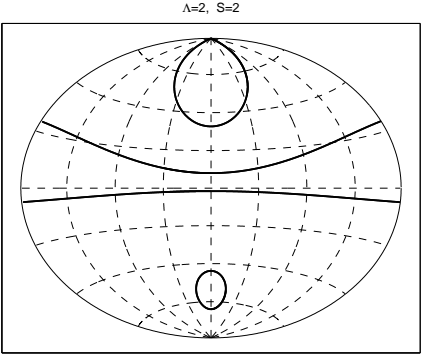} }
 \caption{Vesicle dynamics on the $\Theta-\Phi$ atlas.}
 \label{figure:stp01}
 \end{figure}

A solution of the equation (\ref{eq_on_rho}) behaves differently at $\Lambda<1$ and at
$\Lambda>1$. At $\Lambda<1$ the variable $\rho$ trends to infinity as time grows, thus
$U_5$ has its utmost value be equal to $\sqrt{1-\Lambda^2}$, that corresponds to a stable
point (tank-treading behavior). At $\Lambda>1$ the variable $\rho$ experiences
oscillations. Then one conclude from Eqs. (\ref{eq_on_rho}-\ref{PhiandTheta}), that the
vesicle evolution is described by a limit cycle, its characteristics are determined by
values of the integrals ${J}$ and $\Upsilon$. If the angle $\Phi$ grows (decreases)
unlimited, the vesicle is in the tumbling regime. On the contrary, if $\Phi$ is bounded,
one deals with the trembling regime. It is convenient to represent vesicle dynamics as a
geographic atlas, where $\Theta$ and $\Phi$ play roles of latitude and longitude,
correspondingly, see Fig. \ref{figure:stp01}. The trembling regime corresponds to a
closed curve on the $\Theta$-$\Phi$ atlas, which does not surround a pole. The tumbling
regime corresponds to a curve separating the poles. In Fig. \ref{figure:stp01} such
curves terminate at the boundaries of the atlas having the same $\Theta$ on the ends,
since points on the right and on the left boundaries of the atlas with the same latitude
are physically identical. For the truncated system (\ref{truncated}) the tumbling and
trembling regimes coexist at any $\Lambda>1$. A choice between the regimes is determined
by a value of $\Upsilon$. If $\cos \Upsilon>2\Lambda/(\Lambda^2+1)$, then the limit cycle
corresponds to tumbling, otherwise the cycle corresponds to trembling. If $\Upsilon$
takes one of its boundary values, that is if $\cos \Upsilon = 1/\Lambda$, then the limit
cycle degenerates into a point
 \begin{equation}
 \Phi=0, \qquad \cos\Theta=1/\Lambda.
 \label{statio}
 \end{equation}
Thus, the truncated system of equations (\ref{truncated}) has two stationary points
corresponding to the tank-treading regime.

\subsection{Strong external flows, slow dynamics}

We demonstrated that the truncated system of equations (\ref{truncated}) can be
completely integrated. However, the system cannot be directly used for an analysis of the
vesicle dynamics in the limit of strong shears $S\gg1$. Indeed, the system leads to a
dependence of its solution on initial conditions and admits different limit cycles for
any $\Lambda>1$. Both these properties contradict obviously to the results obtained in
Subsec. \ref{subsec:general}. The contradiction is resolved if one restores the terms
omitted in the truncated system (\ref{truncated}) and originating from the last
(non-linear in $U$) term in the right-hand side of Eq. (\ref{diless}). The restored terms
provide a relatively slow evolution of both integrals of motion, ${J}$ and $\Upsilon$,
which leads to a well defined behavior, independent of the initial conditions.

Our nearest goal is to deduce the equations of motion for ${J}$ and $\Upsilon$. We
consider the case $\Lambda>1$ where the truncated system of equations (\ref{truncated})
leads to limit cycles. Then, analyzing the complete system of equations, one can separate
fast motion along the limit cycles and relatively slow evolution of the integrals of
motion on scales larger than the cycle period. Note, that a typical time of the fast
dynamics is $\tau_\ast/S$ whereas a typical time of the slow dynamics is $\tau_\ast$, the
large ratio of the times justifies the separation. The equation controlling the slow
evolution can be found by averaging over the cycle period of the expressions for the time
derivatives of ${J}$ and $\Upsilon$ obtained from the complete system (\ref{diless}). At
this averaging, one can use the fast dynamics described by the system (\ref{truncated}).
The result can be schematically represented in the form $\partial_t \Upsilon = \dot
\Upsilon(\Upsilon,J,\Lambda)$, $\partial_t{J} = \dot{J}(\Upsilon, {J},\Lambda)$. Explicit
expressions for $\dot\Upsilon$ and $\dot{J}$ are quite cumbersome, so we do not present
here its final form.

One can check, that the ultimate value of ${J}$ is equal to zero at any $\Lambda$. That
is why below we consider the case ${J}=0$, and find the slow dynamics for the quantity
$\Upsilon$ at the condition. Then $\dot\Upsilon$ is a function of $\Lambda$ and
$\Upsilon$, the function can be found from Eqs.
(\ref{eq_on_Theta},\ref{eq_on_Phi},\ref{second})
 \begin{equation}
 \dot\Upsilon=\left\langle(\partial \Upsilon/\partial \Theta)
 \cos(3\Theta) \right\rangle ,
 \label{sloweq}
 \end{equation}
where angular brackets mean averaging over the cycle period. The evolution of $\Upsilon$
described by the equation $\partial_t\Upsilon=\dot\Upsilon$ leads to a stationary value
which can be found from the equation $\dot\Upsilon=0$. The value corresponds to a limit
cycle, which is stable provided $\partial \dot \Upsilon/\partial \Upsilon <0$. In
accordance with the analysis, made in the previous subsection, a character of the limit
cycle depends on the value of $\Upsilon$. If $\cos \Upsilon>2\Lambda/(\Lambda^2+1)$, then
the limit cycle corresponds to tumbling, otherwise the cycle corresponds to trembling. A
numerical investigation based on Eq. (\ref{sloweq}) shows that the boundary value of
$\Upsilon$, $\Upsilon=\arccos[2\Lambda/(\Lambda^2+1)]$, is achieved at $\Lambda\approx
1.52$, larger $\Lambda$ correspond to tumbling, smaller $\Lambda$ correspond to
trembling.

At $\Lambda<\sqrt2$ all the limit cycles appear to be unstable. That means that the
condition the stationary point (\ref{statio}) appears to be an attractor. Thus, the
system falls to the stationary point that corresponds to the tank-treading regime. Note
that the position of the stationary point is slightly shifted due to the presence of the
additional term in the equation for $\Theta$ (\ref{eq_on_Theta}). The new position is
 \begin{equation}
 \Theta=\arccos(1/\Lambda), \qquad
 \Phi=\frac{\cos(3\Theta)}{2S\sin\Theta},
 \label{statio2}
 \end{equation}
in the main approximation in $1/S$. A correction to the value $\Theta=
\arccos(1/\Lambda)$ is of the order $1/S^2$.

\subsection{Extremely strong flows}

The above analysis is, strictly speaking, correct for flows characterized by $S<
1/\sqrt\Delta$. For stronger flows, $1/\sqrt\Delta\lesssim S$, our consideration should
be extended. Some additional terms of the higher order in $u$ should be taken into
account in the equation for $u$, which for $1/\sqrt\Delta\lesssim S$ are larger than
those kept in Eq. (\ref{main1}). Leading terms of such kind can be estimated as $su$,
they originate, say, from account of deviations of the vesicle shape from a spherical
one. The terms are associated with the contribution $\bm v^{(s)}$ to the membrane
velocity, whereas the second term in the right hand side of Eq. (\ref{main1}) is
associated with the contribution $\bm v^{(\kappa)}$ to the membrane velocity, see Eq.
(\ref{decompos}).

However, there is an essential difference between the terms $\bm v^{(s)}$ and $\bm
v^{(\kappa)}$. If one omits the term $\bm v^{(\kappa)}$ then there appears a symmetry of
the vesicle dynamics expressed in terms of the equation for $u$, which is invariant under
simultaneous time and space invertions, see Subsection \ref{subsec:very}. Since the axes
of our reference system are attached to the eigen vectors of the strain matrix $\hat s$,
the space inversion is equivalent to the transformation $\varphi\to-\varphi$,
$\theta\to\theta$, that is it can be written as $\Phi\to-\Phi$ and $\Psi\to-\Psi$ in
terms of the ``angles'' (\ref{U_representation}). Therefore the equations for $\Theta$
and $\Phi$ should be invariant under the transformation
 \begin{equation}
 t\to-t, \quad \Phi\to-\Phi, \quad \Theta\to\Theta.
 \label{symmetry}
 \end{equation}
One can easily check that the truncated equations (\ref{truncated}) are invariant under
the transformation (\ref{symmetry}). However, our analysis demonstrated that the symmetry
survives even though higher in $u$ terms will be taken into account provided
$\kappa\to0$.

If small in $u$ corrections to the truncated equation (\ref{truncated}) will be taken
into account then the system of limit cycles, characteristic of this equation at
$\Lambda>1$, will survive with small perturbations. Next, if the term $\bm v^{(\kappa)}$
in the expression (\ref{decompos}) is neglected all the limit cycles will be neutral
(they are no stable, no unstable). Indeed, the cycle passing through a point
$\Phi=0,\Theta$ cannot be stable, since then due to the symmetry (\ref{symmetry}) it
should remain stable after the time inversion. Thus, the corrections to the truncated
equation (\ref{truncated}) keep its main property which is an existence of an additional
integral of motion (leading to the neutrality of the limit cycles).

Thus only the term $\bm v^{(\kappa)}$ proportional to $\kappa$ can destroy the
integrability corresponding to the neutral limit cycles. It produces a selection leading
to stable limit cycles or to stable stationary points. Since the selection is produced
among limit cycles slightly disturbed in comparison with the ones corresponding to the
truncated equation (\ref{truncated}), the results will be the same in the main
approximation in $\Delta$. Thus the above results obtained for strong flows can be
immediately extended to the extremely strong flows, where $S>1/\sqrt\Delta$.

\section{Conclusion}
\label{sec:conclu}

We have investigated dynamics of nearly spherical vesicles in an external stationary
flow, being mainly focused on the shear flow. The general calculational scheme developed
for nearly spherical vesicles enabled us to analyze the dynamics in detail. The scheme is
based on solving the $3d$ hydrodynamic (Stokes) equations with boundary conditions posed
on the membrane. Besides the membrane bending elasticity we have taken into account the
internal membrane viscosity, leading to an additional dissipation mechanism.

There are essentially different regimes of the vesicle dynamics dependent on a relation
between the strain of the external flow and the vesicle relaxation rate. In the weak
flows the vesicle shape is close to an equilibrium one which is a prolate ellipsoid. Then
a role of the external flow is reduced mainly to an orientation of the ellipsoid. In the
strong flows the vesicle shape and orientation are determined by the flow. We established
the vesicle behavior for different strengthes of the external flow including its
orientation relative to the external velocity.

The most interesting phenomenon in the vesicle dynamics is transition to tumbling regime
which occurs at increasing the generalized viscous contrast (\ref{defa}) or at increasing
the rotational component of the external flow. We demonstrated that in weak flows a
direct transition from tank-treading to tumbling occurs whereas in strong flows an
intermediate regime, trembling, is realized. The behavior is in accordance with
experiment of Kantsler and Steinberg \cite{KS06} and corresponds qualitatively to
numerics of Noguchi and Gompper \cite{NG07}. The phase diagram we obtained is plotted in
Fig. \ref{figure:phase_diagram} where the variables $S$ and $\Lambda$ are some
combinations of observable quantities (\ref{selfsim}).

The possibility of an intermediate regime between tank-treading and tumbling was
discussed theoretically, first by Misbah \cite{M06} and then (qualitatively) by Noguchi
and Gompper \cite{NG07}. Note, however, that the calculational scheme used by Misbah
\cite{M06} and then by Vlahovska and Gracia \cite{VG07} is not self-consistent though
formally the authors have taken into account principal terms of the equation for the
vesicle distortions. The situation cannot be improved by introducing higher order terms
related to the external flow. And only after introducing higher order terms related to
the membrane bending elasticity into the equation it enables one to find, say, boundaries
of the region where the trembling regime is realized.

The transitions tank-treading to tumbling and tank-treading to trembling have essentially
different scenarios. The first transition is described as a saddle-node bifurcation
whereas the second one is described as a Hopf bifurcation. Our theory predicts existence
of a special point on the diagram where the two above transition lines merge. One expects
an essential ``critical'' slowness of the vesicle dynamics near the point. Therefore it
plays a role analogous to some extent to the critical point on fluid phase diagram.

Near the transitions thermal fluctuations are relevant which smear the transitions. One
expects that the effect is especially strong near the special point on the phase diagram.
Role of the thermal fluctuations, which can be examined in the spirit of the works
\cite{96KLM,05CKLT,07Tur}, constitutes a subject of special investigation, to be done
separately. Here we note only, that due to fluctuations one should be careful in
comparison an experiment with the phase diagram obtained in our work since the diagram is
deduced ignoring the fluctuations.

It is worth to mention the effect, related to the thermal fluctuations, recently
discovered by Kantsler et. al. \cite{07KSS}. It was shown that the relaxational dynamics
of a vesicle in external elongational flow is accompanied by the formation of wrinkles on
a membrane. Theoretical investigation of this effect presented in \cite{07TV} was based
on the theory, developed in this paper. It was shown that the formation of wrinkles is
related  to the dynamical instability induced by negative surface tension of the
membrane.

We investigated nearly spherical vesicles assuming that the excess area factor $\Delta$
is small. It enables one to formulate a powerful calculational scheme enabling to find
details of the vesicle dynamics analytically. We have doubts that the scheme can be
generalized for a general case $\Delta\sim1$. Probably, the case can be investigated only
numerically. However, the general approach should be the same: one has to solve the
stationary Stokes equation inside and outside the vesicle with the boundary conditions
(\ref{sincomp},\ref{normal},\ref{tangent}) at a given vesicle shape and then to use the
equation (\ref{kinema}) to formulate an equation for membrane distortions.

\acknowledgments

We wish to thank V. Kantsler, I. Kolokolov, and V. Steinberg for numerous valuable
discussions. This work has been partially supported by RFBR grant 06-02-17408-a and joint
RFBR-Israel grant 06-02-72028. KT and SV acknowledge the financial support from
``Dynasty'' and RSSF foundations.

\appendix

\section{General dynamics}
\label{sec:dynamics}

In the appendix we present the derivation of Equation (\ref{equ1}). For brevity we
introduce viscosity contrast parameter $\lambda = \tilde\eta/\eta$ and surface viscosity
parameter $\mu = \zeta/\eta r_0$

At low Reynolds numbers velocity field of fluid is completely determined by the boundary
conditions. One can solve the Stokes equation (\ref{Stokes}) inside and outside the
vesicle membrane with boundary conditions on the membrane and a fixed boundary condition
at the infinity (\ref{graddec}). This allows one to exclude the fluid motion and arrive
to the close equation onto membrane surface. On the way one obtains integral equation,
describing membrane surface motion \cite{BRSBM04, 87Lad}. Further analytical progress
is possible only at the limit $\Delta\ll1$, when it is possible to expand the equation
describing the membrane dynamics in series over $\Delta$. This expansion has no regular
limit  at $\Delta\rightarrow 0$. Physically, exactly spherical vesicle can not conserve
its volume and area in arbitrary small external flow: to describe the vesicle behavior in
an external flow one should restore final surface compressibility of the vesicle
membrane. As a consequence, expansion series of some quantities, which depend on the
vesicle shape, starts from zero or even negative order in $\Delta$ terms.

First derivation of this type of equation was presented in \cite{S99} in particular case
of equal inner and outer viscosities, $\eta=\tilde\eta$ and zero surface viscosity,
$\zeta = 0$ (see also further generalizations \cite{M06,VG07}, where case
$\lambda\neq1$ was considered). To proceed to the derivation, let us first exclude fluid
motion inside and outside the vesicle. We parameterize velocity field inside and
perturbed part of velocity field ${\bm v}-{\bm V}$  outside the vesicle by the functions
\begin{eqnarray}\label{XYZin_XYZout}
    X^\mathrm{in}, Y^\mathrm{in}, Z^\mathrm{in}
    \quad
    \mathrm{and}
    \quad
    X^\mathrm{out}, Y^\mathrm{out}, Z^\mathrm{out}
\end{eqnarray}
correspondingly. The functions (\ref{XYZin_XYZout}) are functions of spherical angles
$\theta$, $\varphi$. Procedure of restoration of velocity field into the bulk from the
functions is known as Lamb solution. The physical meaning of the functions
(\ref{XYZin_XYZout}) is the following. On a sphere with the radius $r_0$ the components
of velocity and vorticity fields are given by
\begin{eqnarray}\label{functions_on_the_sphere}
    v^r = X^\mathrm{in}, \quad r_0\partial_rv^r = Y^\mathrm{in},\quad
    r_0\omega^r = Z^\mathrm{in},
\end{eqnarray}
where $\bm{\omega}$ is vorticity. Using the  Lamb solution one can make analytical
continuation of the velocity into whole interior of the vesicle. We expand each of
functions (\ref{XYZin_XYZout}) into series over spherical harmonics: for example,
\begin{eqnarray}\label{n_expansion}
    X^\mathrm{in}
    =
    \displaystyle\sum\limits_{l\geq0,m}{X^\mathrm{in}}^{l,m} {\cal Y}_{l,m}(\theta,\phi),
\end{eqnarray}
where $m$ runs from $-l$ to $l$. Assuming that the velocity field is a regular function
inside the sphere, one finds by solving of (\ref{Stokes}) that
\begin{eqnarray}\label{v_representation}
    \bm v
    =
    \mathrm{grad}\,\Pi + \mathrm{rot} [\bm r \varpi_1]
    +
    \mathrm{rot}\,\mathrm{rot}[\bm r \varpi_2]
\end{eqnarray}
where
\begin{eqnarray}&\label{v_grad}
    \Pi
    =
    \displaystyle\sum\limits_{l}
    \displaystyle\frac{r^{l}}{(r_0)^{l-1}}
    \displaystyle\frac{(l+1)X^{\mathrm{in},l}-Y^{\mathrm{in},l}}{2l}
\end{eqnarray}
\begin{eqnarray}\label{v_rot}
    \varpi_1
    =
    -\sum\limits_l
    \frac{r^l}{(r_0)^{l}}
    \displaystyle\frac{Z^{\mathrm{in},l}}{l(l+1)}
\end{eqnarray}
\begin{eqnarray}\label{v_rotrot}
    \varpi_2
    =
    \sum\limits_l
    \displaystyle\frac{r^{l+2}}{(r_0)^{l+1}}
    \displaystyle\frac{Y^{\mathrm{in},l}-(l-1)X^{\mathrm{in},l}}{2l(l+1)}.
\end{eqnarray}
In (\ref{v_grad}-\ref{v_rotrot})  the summation goes over $m$ as well, although we omit
it in the equatoins for brevity. The pressure field
\begin{eqnarray}\nonumber
    P = \tilde\eta\sum\limits_l
    \displaystyle\frac{r^l}{\left( r_0 \right)^{l+1}}
    \displaystyle\frac{(2l+3)\left(Y^{\mathrm{in},l} - (l-1)X^{\mathrm{in},l}\right)}{l}
\end{eqnarray}
is induced only by the last velocity component in (\ref{v_representation}).

\subsection{Curvilinear coordinates}

In the Appendix we use the curvilinear coordinates for the vector quantities in contrast
to the main text, where we used cartesian coordinates for clarity. Let us parametrize the
vesicle surface by two internal coordinates $\xi^{1,2}$. Leaving indices $i,j,k,\ldots$
for vector quantities, written in cartesian coordinates, we identify vector quantities,
projected onto the membrane surface and written in the internal coordinates by indices
$\alpha,\beta,\ldots$. We denote the metric and curvature tensors as $g_{\alpha\beta}$
and $h_{\alpha\beta}$ correspondingly. Recall, that by definition the mean curvature of
the membrane is given by $H = -g^{\alpha\beta}h_{\alpha\beta}$, where as usual we assume
summation over repeating indices. Also we introduce covariant derivative on the surface
$\nabla_\alpha^{\!\scriptscriptstyle \perp}$, and for any quantity, for example $\sigma$,
we use the notation $\nabla_\alpha^{\!\scriptscriptstyle\perp}\sigma = \sigma_{;\alpha}$.

Force balance equations (\ref{normal},\ref{tangent}), written in the curvilinear
coordinates, have the following form:
\begin{eqnarray}&\hskip-20pt \nonumber
    -\zeta h^{\alpha\beta}\mathbb{S}_{\alpha\beta}\big\vert_\mathrm{m}
    +
    H\sigma
    -
    \kappa\left[H(H^2/2 - 2K) + \Delta^{\!\scriptscriptstyle \perp}H
    \right] =
    \\[5pt]& \hskip50pt
    =
    \lfloor P\rfloor\big\vert_\mathrm{m},
\end{eqnarray}
\begin{eqnarray}\hskip-10pt\label{tangent_curvilinear}
    \nabla^{\!\scriptscriptstyle \perp}_\alpha\sigma
    +
    \zeta g^{\gamma\beta}\nabla^{\!\scriptscriptstyle \perp}_\beta
       \mathbb{S}_{\alpha\beta}\big\vert_\mathrm{m}
    =
    l_i\left\lfloor
    \eta\mathbb{S}_{i\alpha}
    \right\rfloor
    \big\vert_\mathrm{m},
\end{eqnarray}
where $\mathbb{S}$ is doubled symmetric part of velocity gradient, in Cartesian
coordinates $\mathbb{S}_{ij} = \partial_iv_j+\partial_jv_i$ and brackets
$\lfloor\ldots\rfloor$ denote the difference  between the ``inside" and  ``outside"
values. The index ``m" near vertical line indicates, that the equations are written on
the membrane surface.

We choose the  two spherical angles as the internal coordinates, see (\ref{function_u}).
While developing a perturbation theory over $\Delta$, it is convenient to use the method
of domain perturbations, and to use the sphere with the radius $r_0$, parameterized by
the spherical angles, as intermediate surface. Each point on the membrane surface
corresponds to a point on the sphere with radius $r_0$, having the same spherical angles.
To distinguish vector quantities and equations, defined on the intermediate surfaces, we
write index ``s" near vertical curve instead of index ``m" for the membrane surfaces, see
for example (\ref{tangent_curvilinear}). Any quantity defined in a bulk or on a membrane
surface can be represented as a taylor expansion through its value and its derivatives
defined on a intermediate sphere. This method is well defined for quasi-spherical
membranes with $\Delta \ll 1$.

For the sphere we introduce one more object -- directional derivative, defined by the
equation
\begin{eqnarray}\label{oblique_derivative}
    \nabla^\ast_\alpha
    =
    g_{\alpha\beta}\frac{\epsilon^{\beta\gamma}}{\sqrt{\mathrm{det}g}}
    \nabla^{\!\scriptscriptstyle \perp}_\gamma\bigg\vert_\mathrm{s},
\end{eqnarray}
where $\epsilon^{\beta\gamma}$ is antisymmetric unit symbol, $\epsilon^{12} = 1$. Also we
denote Beltrami-Laplace operator on the sphere as $\hat{\cal O}/r_0^2$, such that
$\hat{\cal O}$ is Beltrami-Laplace operator on unit sphere.

\subsection{Utilization of Lamb solution.}

In this subsection we express a list of quantities, which are involved in process of
satisfying of boundary conditions, through functions (\ref{XYZin_XYZout}). We use the
following notations. Suppose function $g$ is a function of spherical angles $\theta$,
$\varphi$ and $\mathrm K(l)$ is some function of angular harmonics $l$. Then linear
operator $\hat{\mathrm K}^+$ is defined as
\begin{eqnarray}\label{K_plus}
    \hat{\mathrm K}^+ g
    =
    \sum_{l>1,m}\mathrm K (l)\ g_{l,m}.
\end{eqnarray}
Also we introduce linear operator $\hat{\mathrm K}^-$, whose definition can be obtained
from (\ref{K_plus}) through replacing $l\rightarrow -(l+1)$. Also we denote
\begin{eqnarray}
    \hat{\mathrm K}
    =
    \hat{\mathrm K}^+ - \hat{\mathrm K}^-,
    \quad
    \hat{\mathrm Q}
    =
    \lambda\hat{\mathrm K}^+ -\hat{\mathrm K}^-.
\end{eqnarray}
We need in pressure, velocity and some components of symmetric part of velocity gradient
$\mathbb{S}_{ij}$ on sphere with radius $r_0$. Here we list all quantities on inner side
of the sphere, omitting index $in$. Quantities on outer side of the sphere can be
obtaining by changing $l\rightarrow-l-1$ and plugging functions (\ref{XYZin_XYZout}) with
index $out$. Pressure inside is
\begin{eqnarray}
    p=
    \text{r}_0\nu[\hat{\mathbb K}^{px+}\ X
    +
    \hat{\mathbb K}^{py+}\ Y]\bigg\vert_\mathrm{s},
\end{eqnarray}
where
\begin{eqnarray}
    {\mathbb K}^{px}(l)=-\frac{(l-1)(2l+3)}{l}
    ,\quad
    {\mathbb K}^{py}(l)=\frac{2l+3}{l}.
\end{eqnarray}
The velocity is given by
\begin{eqnarray}
    v_\alpha
    =
    -\nabla^{\!\scriptscriptstyle\perp}_\alpha
    \hat\mathcal{O}^{-1}(2X+Y)
    -\nabla^\ast_\alpha\hat\mathcal{O}^{-1}Z.
\end{eqnarray}
Tangential-radial components of velocity strain $\hat\mathbb{S}$ is
\begin{eqnarray}\nonumber
    \hskip-10pt
    \mathbb{S}_{r\alpha}
    =
    \bigg\{
    \nabla_\alpha^{\!\scriptscriptstyle \perp}
    \bigg[\hskip-10pt&&
    \hat{\mathbb K}^{sx+}_{r\tau}\,X
    +
    \hat{\mathbb K}^{sy+}_{r\tau}\,Y
    \bigg]
    -
    \nabla_\alpha^\ast
    \hat{\mathbb K}^{s z+}_{r\tau}\,Z
    \bigg\}
    \bigg\vert_\mathrm{s}
    \label{varsigma_r_tau}.
\end{eqnarray}
In (\ref{varsigma_r_tau}) necessary for us coefficients $\hat{\mathbb K}$ are
\begin{eqnarray}
    {\mathbb K}^{sx}_{r\tau}(l)
    =
    \frac{l-1}{l(l+1)}, \quad
    {\mathbb K}^{sy}_{r\tau}(l)
    =
    \frac{2l+1}{l(l+1)}.
\end{eqnarray}
Tangential-tangential components of $\hat\mathbb{S}$ are
\begin{eqnarray}&\label{S_tau-tau}\hskip-40pt
    \mathbb{S}_{\alpha\beta}
    =
    \bigg\{
    2Xg_{\alpha\beta} -
    \left[\hat{\cal O}^{-1}
    \left[4X+2Y\right]
    \right]_{;\alpha\beta}
    -\\ \nonumber &\hskip40pt -
    \left[
    \nabla^\ast_\alpha\nabla^{\!\scriptscriptstyle \perp}_\beta
    +
    \nabla^\ast_\beta\nabla^{\!\scriptscriptstyle \perp}_\alpha
    \right]
    \left[\hat{\cal O}^{-1}Z\right]
    \bigg\}
    \bigg\vert_\mathrm{s}.
\end{eqnarray}
In the interesting case of $Y=Z=0$ one obtains from (\ref{S_tau-tau})
\begin{eqnarray}\label{S_tau_tau_1}
    g^{\beta\gamma}\nabla^{\!\scriptscriptstyle\perp}_\gamma\mathbb{S}_{\alpha\beta}
    \bigg\vert_\mathrm{s}
    =
    -2\nabla^{\!\scriptscriptstyle \perp}_\alpha
    \left[1+2\hat{\cal O}^{-1}\right]X\bigg\vert_\mathrm{s}
\end{eqnarray}
and
\begin{eqnarray}\label{S_tau_tau_2}
    h^{\alpha\beta}\mathbb{S}_{\alpha\beta}\big\vert_\mathrm{s} = 0.
\end{eqnarray}
The radial derivative is given by
\begin{eqnarray}
    \partial_r\mathbb{S}_{rr} = 2\partial_r^2v^r
    =
    \hat\mathbb{K}^{sx+}_{rrr}X + \ldots,
\end{eqnarray}
where we kept the only necessary contribution from the unction $X$ with
\begin{eqnarray}
    \mathbb{K}^{sx}_{rrr}(l) = 2(1-l^2).
\end{eqnarray}

\subsection{Obtaining of the evolution equation in main approximation in $\Delta$.}
\label{subsec:appendix:zero_approximation}

In this Subsection we return to the derivation of Eq. (\ref{equ1}). We have seven unknown
scalar functions, six in (\ref{XYZin_XYZout}) and surface tension $\sigma$, which depend
on spherical angles $\theta$, $\varphi$. To find these functions, one should satisfy all
the
 boundary conditions on the membrane. Three boundary
conditions come from velocity continuity in the whole space. For us it is convenient to
use three equivalent conditions: continuity of normal component of velocity $l^iv^i$,
continuity of normal derivative of normal component of velocity $l^il^k\partial_kv^i$ and
normal component of vorticity $l^k\omega^k$. Another three conditions come from the
continuity of the momentum flux, (\ref{normal},\ref{tangent}). Seventh condition
corresponds to the surface flow incompressibility, see (\ref{sincomp}). After the seven
scalar filed are found, one should use the relation between the velocity and the temporal
derivative of function $u$ (\ref{function_u}) presented at (\ref{kinema}) and obtain
required equation of motion. In this section we put $\kappa=0$, and do not take into
account the terms arising from the bending force of the membrane.

To obtain the dynamical equation with the accuracy up to $n$-th order of $\sqrt{\Delta}$,
one should satisfy all boundary condition with the same accuracy. It is convenient to
find the next in $\sqrt{\Delta}$ correction to the equation using the recursive
procedure. In accordance with the scheme we represent any quantity, for example
$X^\mathrm{in}$, as a series
\begin{eqnarray}
    X^\mathrm{in} = \sum\limits_{n=0}^\infty X^\mathrm{in}_n,
\end{eqnarray}
where $X^\mathrm{in}_n\propto \Delta^{n/2}$. For vector quantities which have low indices
we put the index $n$ on the left from the main letter to avoid mixing of different
indices.

We represent the local value of the surface tension as $\sigma = \bar\sigma +
\tilde\sigma$, where $\bar\sigma$ does not depend on angles $\theta,\varphi$ and the
integral of $\tilde\sigma$ over the angles is zero. The reason of the division is
different scaling laws of parts $\bar\sigma$ and $\tilde\sigma$ with $\Delta$: expansion
of the quantities in series over $\Delta$ are
\begin{eqnarray}
    \bar\sigma = \sum\limits_{n=-1}\bar\sigma_n, \quad
    \tilde\sigma = \sum\limits_{n=0}\tilde\sigma_n,
\end{eqnarray}
where lower index $n$ near a term corresponds to the term scaling as $\Delta^{n/2}$.

Let us find equation of motion in zero order in $\sqrt{\Delta}$. Surface
incompressibility, i.e. continuity of $l^iv^i$ reads
\begin{eqnarray}\label{v_n_0}
    X^\mathrm{in}_0 = X^\mathrm{out}_0 + X^\mathrm{ext}
\end{eqnarray}
where $X^\mathrm{ext} = h$, definition of $h$ is
\begin{eqnarray}\label{h_function}
    h = s_{ij}l^il^j\big\vert_\mathrm{s}.
\end{eqnarray}
Continuity of $l^il^k\partial_kv^i$ leads to
\begin{eqnarray}\label{d_n_v_n_0}
    Y^\mathrm{in}_0=Y^\mathrm{out}_0+Y^\mathrm{ext},
\end{eqnarray}
where again $Y^\mathrm{ext} = h$. For continuity of normal component of vorticity one
should assume
\begin{eqnarray}\label{w_n_0}
    Z^\mathrm{in}_0 = Z^\mathrm{out}_0 + Z^\mathrm{ext},
\end{eqnarray}
where $Z^\mathrm{ext} = -r_0\epsilon^{ijk}n^j\varsigma^{jk}$ and $\bm n$ is unit vector
on sphere having radius $r_0$. Membrane incompressibility condition yields the equation
\begin{eqnarray}\label{incompressibility_0}
    Y^\mathrm{in}_0 = 0.
\end{eqnarray}

Relation between the velocity field and the dynamics of the vesicle shape has the form
\begin{eqnarray}\label{dot_u_0}
    X_0^\mathrm{in} = \dot u.
\end{eqnarray}

In order to derive the dynamic boundary conditions (\ref{normal}) and (\ref{tangent}), we
should use the expansion of local mean $H$ and Gaussian $K$ curvatures in
$\sqrt{\Delta}$:
\begin{eqnarray}&
    \hskip-17pt
    H_0 = \displaystyle\frac{2}{r_0},
    \quad
    H_1\! =\! -\displaystyle\frac{(2+\hat\mathcal{O})u}{r_0},
    \quad
    H_2 =
    \\[7pt] &
    \ \
    K_0=\displaystyle\frac{1}{r_0^2},
    \
    K_1=\displaystyle\frac{H_1}{r_0}.
\end{eqnarray}

Now we have determined all the objects required to write down (\ref{normal}) and
(\ref{tangent}) in the main approximation over $\Delta$
\begin{eqnarray}&\label{pressure_0}
    -h^{\alpha\beta}\!\ _{0}\mathbb{S}_{\alpha\beta}\big\vert_\mathrm{s}+
    \bar\sigma_{-1}
    H_1+\tilde\sigma_0 H_0
    =
    \lfloor P_0\rfloor,
    \\&\label{tangent_0}
    \zeta\, g^{\beta\gamma}\
    \nabla^{\!\scriptscriptstyle \perp}_\gamma\,
    \!\ _{0}\mathbb{S}_{\alpha\beta}
    \big\vert_\mathrm{s}
    +
    \nabla^{\!\scriptscriptstyle\perp}_\alpha\tilde\sigma_0
    =
    \lfloor\eta\,\!\ _{0}\mathbb{S}_{r\alpha}\rfloor
    \big\vert_\mathrm{s}.
\end{eqnarray}
Values of first terms in (\ref{pressure_0}) and (\ref{tangent_0}) should be taken from
(\ref{S_tau_tau_2}) and (\ref{S_tau_tau_1}) correspondingly. In the approximation
$Z_0^\mathrm{out} =0$, it is convenient to obtain the statement using the directional
divergence (\ref{oblique_derivative}) of Eq. (\ref{tangent_0}): the resulting equation is
a linear homogeneous equation on $Z_0^\mathrm{out}$. Taking simple divergence of Eq.
(\ref{tangent_0}), one can find alternating part of the surface tension
\begin{eqnarray}\label{tilde_sigma_0}
    \frac{1}{\eta r_0^2}\,\tilde\sigma_0 =
    \left\{2\mu\left(1+2{\cal O}^{-1}\right)
    +
    \hat{\mathbb Q}^{sx}_{r\tau}
    \right\}
    \dot u
    -\frac{5}{2}h.
\end{eqnarray}
Substituting (\ref{tilde_sigma_0}) it into (\ref{pressure_0}) one eventually obtains:
\begin{eqnarray}&\label{equation_of_motion_0}
    \dot u_0
    =
    \hat a^{-1}\left[
    \displaystyle\frac{\bar\sigma_{-1}}{\eta r_0^2}\,(\hat{\cal O}+2)\,u+10h
    \right],
    \\[7pt] &\nonumber
    \hat a
    =
    2\hat{\mathbb Q}^{sx}_{r\tau}
    -\hat{\mathbb Q}^{px}
    +
    4\mu\left(1+2{\cal O}^{-1}\right),
    \\ &
    a(l)
    =
    \frac{2 (\lambda +1) l^3+3 (\lambda +1) l^2-5 \lambda+4}{l (l+1)}
    + 4\mu\frac{l^2+l-2}{l(l+1)}.
\end{eqnarray}
Surface tension $\bar\sigma_{-1}$ can be obtained from the membrane area conservation law
in the main approximation, $\Delta_2(u)=0$. Formally, $\dot u_0$ in
(\ref{equation_of_motion_0}) is order of $1$, that is reflected by low index ``0". Note,
that rotational part of external flow $\omega$ does not entered in
(\ref{equation_of_motion_0}). This is the consequence of the fact, that the rotational
part of the external flow enters in the exact equation on $u$ though extension of time
derivative, $\partial_t\rightarrow(\partial_t-\omega\partial_\varphi)$, that is
corrections from $\omega$ to (\ref{equation_of_motion_0}) are formally order of
$\sqrt{\Delta}$.

\subsection{Corrections order of $s\sqrt{\Delta}$.}
\label{subsec:s_sqrt_Delta}

In this subsection we find the first corrections in $\sqrt{\Delta}$ to the equation of
motion (\ref{equation_of_motion_0}). One should repeat the steps
(\ref{v_n_0},\ref{d_n_v_n_0},\ref{incompressibility_0},
\ref{dot_u_0},\ref{pressure_0},\ref{tangent_0}), accounting next order in $\sqrt{\Delta}$
and find correction $\dot u_1$ order of $\sqrt{\Delta}$ to $\dot u$
(\ref{equation_of_motion_0}).

In the main approximation we required (\ref{v_n_0}), that
$\left[l^iv^i\right]^\mathrm{out}_0\big\vert_\mathrm{s} =
\left[l^iv^i\right]^\mathrm{in}_0\big\vert_\mathrm{s}$ on the sphere with radius $r_0$.
Here low index ``0" stands for contribution from $X_0,\ldots$, see definition
(\ref{n_expansion}), whereas low index ``1" in (\ref{l_iv_i_1}) corresponds to
contribution from $X_1,\ldots$. Quantity $l^iv^i$ calculated on the membrane surface
through found coefficients $X_0,\ldots$, differs from that on the sphere:
\begin{eqnarray}
    \delta \left[l^iv^i\right]_0\big\vert_\mathrm{m}
    =
    \sum\limits_{n=0} \delta^n \left[l^iv^i\right]_0\big\vert_\mathrm{m}
\end{eqnarray}
where summation index $n$ corresponds to contribution order of $\Delta^{n/2}$. Thus
continuity of normal velocity component $l^iv^i$ in first order in $\sqrt{\Delta}$ looks
like
\begin{eqnarray}&\hskip-70pt\nonumber\label{l_iv_i_1}
    \left[l^iv^i\right]_1^\mathrm{out}\bigg\vert_\mathrm{s} +
    \delta^1 \left[l^iv^i\right]_0^\mathrm{out}\bigg\vert_\mathrm{m}= \\ &
    \hskip40pt=
    \left[l^iv^i\right]_1^\mathrm{in}\bigg\vert_\mathrm{s} +
    \delta^1 \left[l^iv^i\right]_0^\mathrm{in}\bigg\vert_\mathrm{m}.\label{d_n_v_n_1_preliminary}
\end{eqnarray}
where indices ``s" and ``m" stands for sphere surface and membrane surface
correspondingly. Condition can be rewritten as
\begin{eqnarray}\label{d_n_v_n_1}
    X_1^\mathrm{out} = X_1^\mathrm{in},
\end{eqnarray}
since $[l^iv^i]_1\big\vert_\mathrm{s} = X_1$ and
\begin{eqnarray}\nonumber
    \delta^1 \left[l^iv^i\right]_0^\mathrm{in}\big\vert_\mathrm{m}
    =
    \delta^1 \left[l^iv^i\right]_0^\mathrm{out}\big\vert_\mathrm{m}
    =
    -u^{;\alpha} \ _{\scriptscriptstyle 0}\!v_\alpha\big\vert_\mathrm{s}.
\end{eqnarray}
Next boundary condition $l^il^k\partial_k v^i = 0$ in the first order of $\sqrt{\Delta}$
reads
\begin{eqnarray}& \hskip-100pt
    Y_1^\mathrm{in} + \delta^1 \left[l^il^k\partial_k v^i\right]_0^\mathrm{in}\bigg\vert_\mathrm{m}
    =
    \\ &\nonumber
    =
    Y_2^\mathrm{out} + \delta^1 \left[l^il^k\partial_k v^i\right]_0^\mathrm{out}\bigg\vert_\mathrm{m}
    =
    0,
\end{eqnarray}
where
\begin{eqnarray}\hskip-10pt\nonumber
    \delta^1 \left[l^il^k\partial_k v^i\right]_0^\mathrm{in}\bigg\vert_\mathrm{m}
    =
    \frac{1}{2}u\,\hat\mathbb{K}^{sx+}_{rrr}\dot u_{\scriptscriptstyle 0}
    -u^{;\alpha} \left[\mathbb{K}_{r\tau}^{sx+}\dot u_{\scriptscriptstyle 0}\right]_{;\alpha}
    \bigg\vert_\mathrm{s},
\end{eqnarray}
and
\begin{eqnarray}\hskip-10pt\nonumber
    \delta^1 \left[l^il^k\partial_k v^i\right]_0^\mathrm{out}\bigg\vert_\mathrm{m}
    =
    \\ \nonumber
    &\hskip-90pt =
    \left\{\frac{1}{2}u\,\hat\mathbb{K}^{sx-}_{rrr}\dot u_{\scriptscriptstyle 0}
    -u^{;\alpha} \left[\mathbb{K}_{r\tau}^{sx-}\dot u_{\scriptscriptstyle 0}\right]_{;\alpha}
    -
    \frac{5}{2}\left[h^{;\alpha} u\right]_{;\alpha}\right\}\bigg\vert_\mathrm{s}.
\end{eqnarray}
Analog of (\ref{dot_u_0}) now is
\begin{eqnarray}\label{dot_u_1}
    \dot u_1
    =
    -\!\ _{\scriptscriptstyle 0\!}v_{\alpha}\ u^{;\alpha}\big\vert_\mathrm{s}
    +
    X_1.
\end{eqnarray}
In (\ref{dot_u_1}) velocity
\begin{eqnarray}
    \!\ _{\scriptscriptstyle 0\!}v_{\alpha} =
    -\omega\delta_{\alpha\varphi}
    - 2\nabla^{\!\scriptscriptstyle\perp}_\alpha\hat\mathcal{O}^{-1}\dot u_{\scriptscriptstyle 0},
\end{eqnarray}
where $\dot u_0$ is from (\ref{equation_of_motion_0}).

Corrections to boundary conditions (\ref{pressure_0},\ref{tangent_0}) are
\begin{eqnarray} &\hskip-70pt\nonumber
    \bar\sigma_{-1}H_2
    +
    (\bar\sigma_0 +\tilde\sigma_0)H_1
    +
    \tilde\sigma_1 H_0
    =
    \\ & \label{pressure_1}\hskip60pt
    =
    \lfloor P_1\rfloor\big\vert_\mathrm{s}
    +
    \lfloor\delta^1P_0\rfloor \big\vert_\mathrm{m},
    \\[5pt]&\hskip-30pt
    \label{tangent_1}
    \nabla^{\!\scriptscriptstyle\perp}_\alpha \tilde\sigma_1
    =
    \left\lfloor
    \eta\,
    \left[
    l^i\mathbb{S}_{i\alpha}
    \right]_1
    \right\rfloor
    \big\vert_\mathrm{s}
    +
    \left\lfloor
    \eta\,\
    \delta^1\!
    \left[l^i\mathbb{S}_{i\alpha}\right]_0
    \right\rfloor
    \big\vert_\mathrm{m}.
\end{eqnarray}
In (\ref{pressure_1})
\begin{eqnarray}&\nonumber
    \lfloor P_1 \rfloor\big\vert_\mathrm{s}=
    \hat{\mathbb Q}^{px}X_1
    +
    \lambda\hat{\mathbb K}^{py+}\,Y^\mathrm{in}_1
    -
    \hat{\mathbb K}^{py-}\,Y^\mathrm{out}_1,
    \\ &\nonumber
    \lfloor\delta^1P_0\rfloor\big\vert_\mathrm{m}
    =
    u\,\hat{\mathbb Q}^{px}_{r} \,\dot u_{\scriptscriptstyle 0} - 15uh.
\end{eqnarray}
In (\ref{tangent_1})
\begin{eqnarray}&\hskip-150pt\label{1_S_i_alpha}
    \left\lfloor\eta\,
    \left[
    l^i\mathbb{S}_{i\alpha}
    \right]_1
    \right\rfloor\bigg\vert_\mathrm{s}
    =
    \\ &\hskip10pt\nonumber
    =
    \left\{\nabla^{\!\scriptscriptstyle\perp}_\alpha\left[\mathrm{Q}^{sx}_{r\tau} X_1 +\mathcal{Y}_{1,r\tau}^s
    \right]
    +
    \nabla^\ast_\alpha\mathcal{Z}_{1,r\tau}^s\right\}\bigg\vert_\mathrm{s},
    \\[7pt] &\nonumber
    \left\{\mathcal{Y},\mathcal{Z}\right\}_{1,r\tau}^s
    =
    \lambda \mathbb{K}^{s\{y,z\}+}_{r\tau}\{Y,Z\}_1^\mathrm{in}
    -
    \mathbb{K}^{s\{y,z\}-}_{r\tau}\{Y,Z\}_1^\mathrm{out},
\end{eqnarray}
and
\begin{eqnarray}&\hskip-30pt\nonumber
    \left\lfloor\eta\
    \delta^1\!\left[l^i\mathbb{S}_{i\alpha}\right]_0
    \right\rfloor\bigg\vert_\mathrm{m}
    =
    \bigg\{
    u\left[\hat{\mathbb Q}^{{\scriptscriptstyle\mathrm U}x}_{r\tau r}\,
    \dot u_{\scriptscriptstyle 0}\right]_{;\alpha}
    +
    10u\,\nabla^{\!\scriptscriptstyle \perp}_\alpha h
    +\\ &
    +
    (\lambda-1)\left[
    4u^{;\beta}\left[{\cal O}^{-1}\dot u_{\scriptscriptstyle 0}\right]_{;\alpha\beta}
    -
    2u_{;\alpha} \dot u_{\scriptscriptstyle 0}
    \right]
    \bigg\}\bigg\vert_\mathrm{s}.
\end{eqnarray}
We do not write out boundary conditions on $Z_1^{\mathrm{in},\mathrm{out}}$, since
$\mathcal{Z}$ can be excluded from (\ref{tangent_1},\ref{1_S_i_alpha}) by taking
divergence. On the way one obtains, that alternating part of surface tension
\begin{eqnarray}\nonumber
    \tilde\sigma_1
    =
    \hat{\mathcal{O}}^{-1}
    \left\{
    \left\lfloor
    \eta\,
    \left[
    l^i\mathbb{S}_{i\alpha}
    \right]_1
    \right\rfloor^{;\alpha}
    \bigg\vert_\mathrm{s}
    +
    \big\lfloor\eta
    \ \delta^1\!
    \left[l^i\mathbb{S}_{i\alpha}
    \right]_0
    \!{\bigg \vert}_\mathrm{m}
    \big\rfloor^{;\alpha}
    \bigg\vert_\mathrm{s}\right\}.
\end{eqnarray}
Now from boundary condition (\ref{pressure_0}) one can extract correction $\dot u_1$ to
$\dot u$ order of $\sqrt{\Delta}$:
\begin{eqnarray}\label{u_dot_1}
    \dot u_1
    =
    \omega\partial_\varphi
    +
    \bar\sigma_0(\hat{\cal O}+2) u
    +
    \bar\sigma_{-1}\mathcal{L}(u)
    +
    \hat\mathcal{M}_h u.
\end{eqnarray}
Full dynamical equation acquires the form
\begin{eqnarray} &\nonumber
    (\partial_t - \omega\partial_\varphi) u
    =
    \frac{\bar\sigma}{\eta r_0^2}
    \left[\frac{(2+\hat{\mathcal{O}})}{\hat a}u + \mathcal{L}(u)\right]
    +
    \\&
    +
    \frac{10h}{a(2)}
    +
    \hat{\mathcal{M}}_h u.
\end{eqnarray}

\subsection{Restriction onto sector $l=2$.}

It follows from evolution equation, written in the main approximation over
$\sqrt{\Delta}$, that at long times all excess area becomes confined in harmonics $l=2$.
Coupling of higher order harmonics with sector $l=2$ occurs only due to nonlinear terms,
concerned in Subsection \ref{subsec:s_sqrt_Delta}. Relative part of excess area, confined
in higher order harmonics is order of $\Delta$. Hence, back influence of higher order
harmonics on sector $l=2$ appears, if one accounts terms order of $s\Delta$, that is the
influence is negligible in our approximation. Thus, to obtain corrections to dynamics in
sector $l=2$ in the approximation, it is sufficient to account only coupling of mode
$l=2$ with itself.

Bilinear function $\mathcal{L}(u)$ becomes in the case
\begin{eqnarray}
    \mathcal{L}(u) = \frac{24(152+63\lambda)}{(32+23\lambda)^2} u^2
\end{eqnarray}
and nonzero elements of linear operator in the basis (\ref{basispsi}) are
\begin{eqnarray}\nonumber
    {\cal M}_{h}^{51}
    =
    \frac{80\sqrt{3}(56-31\lambda)}{7(32+23\lambda)^2},
    \quad
    {\cal M}_{h}^{15}
    =
    \frac{400(56-41\lambda)}{7\sqrt{3}(32+23\lambda)^2},
\end{eqnarray}
\begin{eqnarray}\nonumber
    {\cal M}_{h}^{23}
    =
    {\cal M}_{h}^{32}
    =
    -\frac{200(14+\lambda)}{7(32+23\lambda)^2}
\end{eqnarray}

\end{document}